\documentclass[preprint]{ptephy_v1}%%%%%% to generate preprint number with ptep logo

\preprintnumber{arXiv 2406.06854} %%% %%% Insert preprint number here
\usepackage{hyperref}
\usepackage{graphicx}% Include figure files
\usepackage{dcolumn}% Align table columns on decimal point
\usepackage{bm}% bold math
\usepackage{upgreek} % up greek font

\begin{document}

%\title{Insert the article title here}
\title{Current-readout technique for ultra-high-rate experiments}

%%%% To generate auto affiliation numbers please use \author{}\affil{} command

%\author{Insert first author name here}
%\affil{Insert first author address here \email{xxxx@xxxx.ac.jp}}

%\author{Insert second author name here}
%\affil{Insert second author address here}

%\author{Insert third author name here}
%\author[3]{Insert fourth author name here} %%% Use optional bracket [3] to change the respective address
%\affil{Insert third author address here}

%\author{Insert last author name here\thanks{These authors contributed equally to this work}}
%\affil{Insert last author address here}

\author[1]{Maki Wakata}
\author[1]{ Shoei Akamatsu}
\author[1]{Takuhiro Fujiie}
\author[1]{Taisei Furuyama}
\author[1]{Lisa Hara}
\author[1]{Yumi Ishikawa}
\author[1]{Tadashi Ito}
\author[1]{Takahiro Kikuchi}

\author[2]{Tsutomu Mibe}

\author[1]{Sachi Ozaki}
\author[1]{Mitsuhiko Yokomizo}

\affil[1]{Department of Physics, Rikkyo University, Tokyo 171-8501, Japan}
\affil[2]{High Energy Accelerator Research Organization (KEK), Ibaraki 305-0801, Japan \email{jiro@rikkyo.ac.jp}}
%\affil[2]{Institute of Particle and Nuclear Studies, High Energy Accelerator Research Organization, Tsukuba, Ibaraki 305-0801, Japan \email{jiro@rikkyo.ac.jp}}

\author[1]{Jiro Murata \thanks{corresponding author} 
}

%\email[corresponding author: ]{jiro@rikkyo.ac.jp}
%\affil[1]{Department of Physics, Rikkyo University, Tokyo 171-8501, Japan \email{jiro@rikkyo.ac.jp} }

%%% To include the collaborator name... Please use the command "\collaborator"
%%% For example: \collaborator{ATLAS Collaboration}

\begin{abstract}
This study developed a new current-readout technique capable of handling measurements with high count rates reaching 1 Gcps. By directly capturing the output current of a photomultiplier as a digitized waveform, we estimated event rates, overcoming the limitations imposed by pulse pileup constraints and deadtimes. This innovative method was applied to a muon spin rotation/relaxation/resonance experiment at the Japan Proton Accelerator Research Complex, demonstrating its expected performance. Furthermore, we explored methods for estimating statistical uncertainty and investigated potential applications in analog-logic OR/AND gates. Our findings reveal that the developed technique opens avenues for developing future non-binary logic circuits operating based on $n$-adic numbers.
\end{abstract}

%\subjectindex{H17, H34, H40}

\maketitle

\section{Introduction}
Handling high-rate counting scenarios with rates of over mega \underline{c}ounts \underline{p}er \underline{s}econd (Mcps) is a crucial technical requirement for conducting experiments at high-intensity accelerator facilities. For instance, a muon beam at the Materials and Life Science Experimental Facility of the Japan Proton Accelerator Research Complex (J-PARC MLF) is expected to have an intensity of $10^{5}\sim 10^6$ particles/beam-pulse \cite{Kadono_2012}. 
Using the muon's lifetime $\tau_\mu = 2.2 \,\upmu$s, the muon decay rate is expected to be $10^6/2.2 \,\upmu{\rm s} \sim \mathcal{O}$ (100G)/s at the maximum.
To exploit such high-intensity beams, a detector may be required to handle a counting rate of 1 Gcps per channel if it covers a solid angle of $\sim$ 1\% of $4\pi$. However, considering that the time scale of a typical detector is $\sim$ 10 ns, counting pulses at over  $1/(10\;{\rm ns})\rightarrow 100\;{\rm Mcps}$ becomes impossible owing to signal pulse pileups. Consequently, fine segmentation aimed at reducing the solid angle of each detector channel and ultra-fast devices aimed at shortening the time scales are usually adopted.

In simple radiation-rate-measuring experiments not requiring event-by-event assessments, identifying individual pulse signals from detectors is unnecessary. Relevant examples of such experiments include simple lifetime and muon spin rotation/relaxation/resonance ($\mu$SR) measurements with negligible backgrounds. 
In this case, the only required information is the positron/electron flux,  $n(t)$ (with $t=0$ denoting the beam-pulse-arrival timing), hitting the detector. 
In this study, we employed digitizer readouts to record the output currents of photomultiplier tubes (PMTs), thus testing their ability to handle a hit rate of  $n\sim 100\;{\rm Mcps}$. Notably, a number of similar waveform digitizing approaches have been adopted \cite{PhysRevD.103.072002,Honda,HAMMAD2019196,AJIMURA2021165742,6544698}. 
While most of these studies aimed to reject pileup events by offline analysis, this study, in contrast, actively records pileup events to enable high-rate measurements.
After developing a system for this current-readout method, we demonstrated the effective application of the developed technique to muon-lifetime and $\mu$SR measurements at the J-PARC.

Furthermore, we also discovered a new technique for accomplishing coincidence measurements between multiple detector channels for background event suppression. Remarkably, the application scope of this technique extends beyond radiation-detection applications to general binary logic circuits, enhancing their information density by utilizing non-binary $n$-adic numbers.

%%%%%%%%%%%%%%%%%%%%%%%%%%%%%%%%%%%%%%%%%%%%%%%%%%%%%%%%%%%%%%%%%%%%%%%%%%%%%%%%%%%%%%
%%%%%%%%%%%%%%%%%%%%%%%%%%%%%%%%%%%%%%%%%%%%%%%%%%%%%%%%%%%%%%%%%%%%%%%%%%%%%%%%%%%%%%

\section{Experiment}
\begin{figure}[htbp]
 \begin{center}
  \includegraphics[width=86mm]{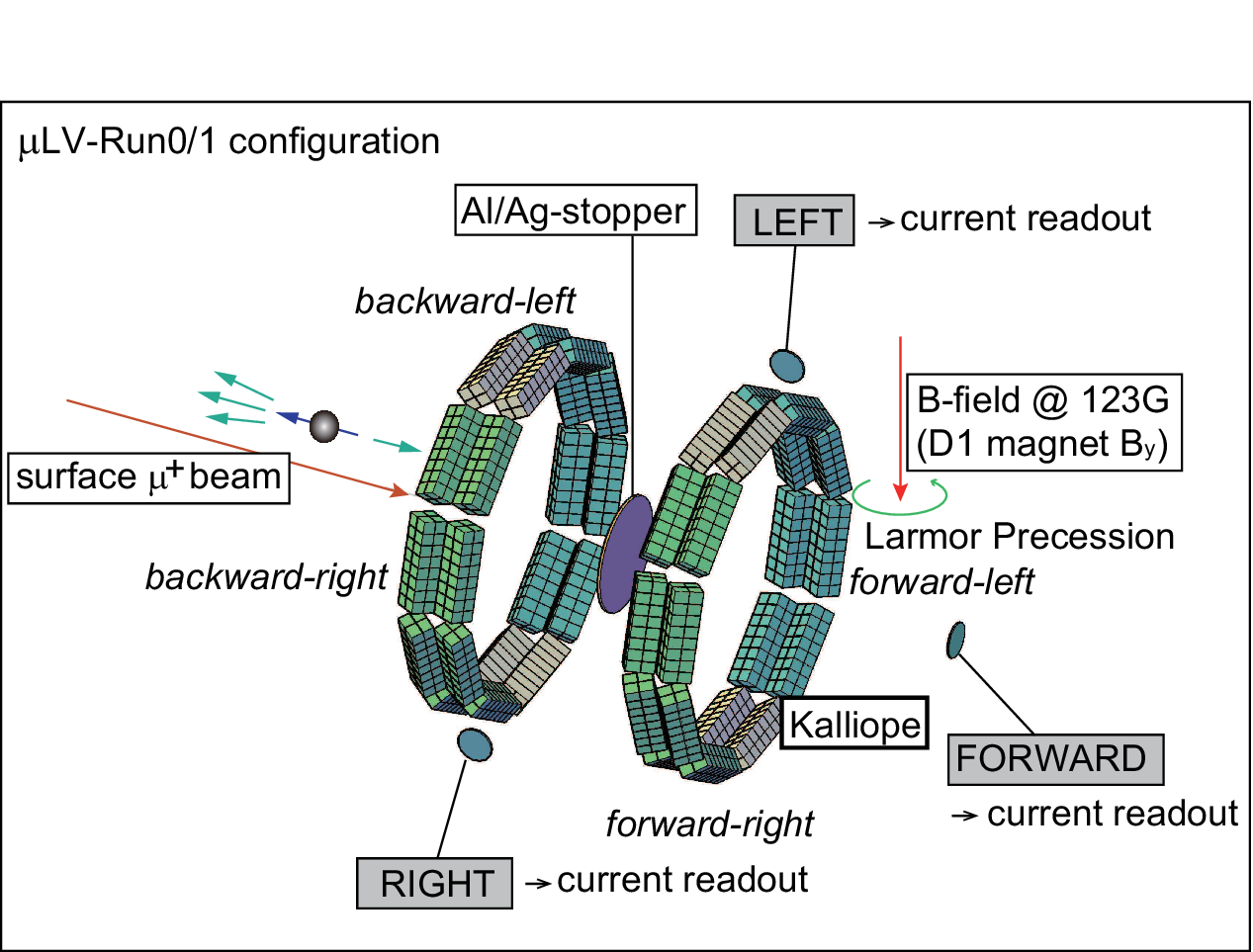} 
 \end{center}
 \caption{Experimental setup of the $\mu$LV-Run0 and Run1 experiments. The framework includes three plastic scintillation counters (LEFT, RIGHT, and FORWARD), along with the Kalliope detector, comprising a 640-ch-inner-plastic-layer and 640-ch-outer-layer \cite{Progress2021, Kojima_2014}.}
 \label{setup}
\end{figure}

We are preparing to execute a new experiment, labeled as the muon Lorentz violation ($\mu$LV) experiment, to perform high-precision muon-lifetime and  $\mu$SR measurements utilizing the pulsed muon beam produced at the J-PARC MLF \cite{Progress2021, Progress2022}. The scientific aim is to compare the shapes of decay spectra, $n(t)$, obtained under varying conditions, without requiring absolute precision in lifetime and spin-precession parameters (such as the Larmor frequency,  $\omega_{\rm L}$ and its parity-violating amplitude $A_{\rm PV}$).
The demonstration experiment involved in this study has been designed to count the number of positrons resulting from
$
\mu^{+}\rightarrow e^{+}+\overline{\nu}_\mu+\nu_e
$
decays that hit single plastic scintillation counters at a counting rate exceeding 100 Mcps, without establishing coincidence pairs. Fig.\ref{setup} presents the experimental setup for the tests, labeled as $\mu$LV-Run0 and Run1, comprising three plastic scintillation counters with PMT readouts. In addition to these, the setup has other plastic scintillation counters finely segmented into 1,280 channels (forming double layers to generate 640 coincidence pairs) with multi-pixel photon counter readouts, known as Kalliope detector  \cite{Kojima_2014}. This Kalliope is a common-use facility permanently set at the beamline, adopted as a reference in this study for comparisons with our system. The primary objective of this study is to overcome the known counting-rate saturation of Kalliope at $\sim 1$ Mcps, attributed to pulse-processing deadtimes and signal pileups.

PMTs with thin plastic scintillators were employed to detect positrons emitted from the decay of positive surface muons stopped within an aluminum sheet with a diameter of 10 cm and thickness of 2 mm (for Run0) or a silver sheet with a thickness of 1mm (for Run1). Specifically, a 3 GeV proton beam was directed at a graphite target to produce pions, which then decayed into muons that stopped at the surface of the production target, creating a ``surface muon beam'' with an energy of 4.1 MeV. This muon beam featured a double-pulse structure comprising two 100-ns-wide pulses separated by a 600-ns-interval, along with a beam repetition rate of $f_{\rm cycle}=1/T_{\rm cycle}=25$ Hz \cite{Kawamura_2014}.

The typical intensity of the surface muon beam could reach as high as  $I_{\rm beam}^{\max}=10^6\; {\rm ppp}$ (\underline{p}articles \underline{p}er double-\underline{p}ulses), with the 3 GeV synchrotron operating at a power of 700 kW. 
The maximum decay rate can be estimated as follows:
\begin{equation}
n_{\rm decay}(t)=\frac{N_{0}}{\tau_\mu} e^{-t/\tau_\mu}\sim 5\times 10^{11} \; /{\rm s} \;\;({\rm at}\;t=0),
\end{equation}
where $N_{0}=I_{\rm beam}^{\max}\times 1$ double-pulse is the muon number contained in a double-beam pulse.
The solid angle covered by a pair of $10\times10\times12$ mm plastic scintillators within the Kalliope setup is $\Omega_{\rm Kalliope}=3.6\times10^{^-4}$. 
Thus, the expected event rate for one Kalliope channel is $n_{\rm decay}(t) \times \Omega_{\rm Kalliope}\sim 180\;{\rm Mcps}$, which exceeds the handling capacity of the electronic equipment within Kalliope. 
To address this, collimators and slits are usually positioned at the beamline to reduce the beam intensity to $1/10 \sim 1/100$ of $I_{\rm beam}^{\max}$. 
However, even higher beam intensities of $I_{\rm beam}\sim 10^7\;{\rm ppp}$ are anticipated at other new beamlines within the MLF, such as the H-line \cite{Yamazaki_2023}. 
This study aims to overcome this bottleneck, thus being able to utilize the maximum beam intensity produced at the J-PARC. 

In this study, measurements were performed using beams with intensities of $I_{\rm beam}^{\rm FULL}=1.9\times 10^5 \; {\rm ppp}$ (``FULL-intensity'' beam) and $I_{\rm beam}^{\rm NORMAL}=3.0 \times 10^4 \;{\rm ppp}$ (``NORMAL-intensity'' beam), obtained by tuning the beamline slits and a lead beam-collimator under $\sim$730 kW power operation of the 3 GeV proton synchrotron. Refer to Appendix \ref{beam} for details on the beam intensity estimation.

The NORMAL-intensity beam is recommended for users performing $\mu$SR experiments at the D1 area to avoid detector saturation effects. For our experiments, three PMTs (with a photocathode featuring a diameter of 25 mm, Hamamatsu H7415  \cite{Hamamatsu}) with thin plastic scintillators (30 mm diameter, BC408, denoted as LEFT, RIGHT, and FORWARD in Fig.\ref{setup}) were employed. The two sideward plastic scintillators (LEFT and RIGHT, with a thickness of 1 mm) were placed with their centers located at a horizontal distance of 360 mm and a vertical downward distance of 76 mm from the center of the stopper (beamline height). Considering that the solid angle of these LEFT or RIGHT counters was $\Omega_{\rm PMT}=4.2\times 10^{-4}$, their maximum expected counting rate for  $I_{\rm beam}^{\rm FULL}$ was
\begin{equation}
n_{\rm LEFT}=n_{\rm RIGHT}=n_{\rm decay} \,\Omega_{\rm PMT}\sim 40\; {\rm Mcps} \; (t=0).
\label{rate_PMT}
\end{equation}
The FORWARD counter (with a thickness of 0.5 mm) was set at a position located 360 mm downstream and 76 mm below the beamline to monitor downstream radiation.

The scintillator planes of the LEFT and RIGHT counters were set parallel to the beam direction to suppress the detection of upstream radiation. 
Conversely, the FORWARD counter was positioned facing the upstream direction to detect positrons from the stopper, as well as other types of radiation traveling in the same direction.
This upstream radiation, more intense than the collimated beam flux by an order of magnitude, is typically undesired in scientific measurements but was utilized in this study for high-intensity tests.

\begin{figure}[htbp]
 \begin{center}
  \includegraphics[width=86mm]{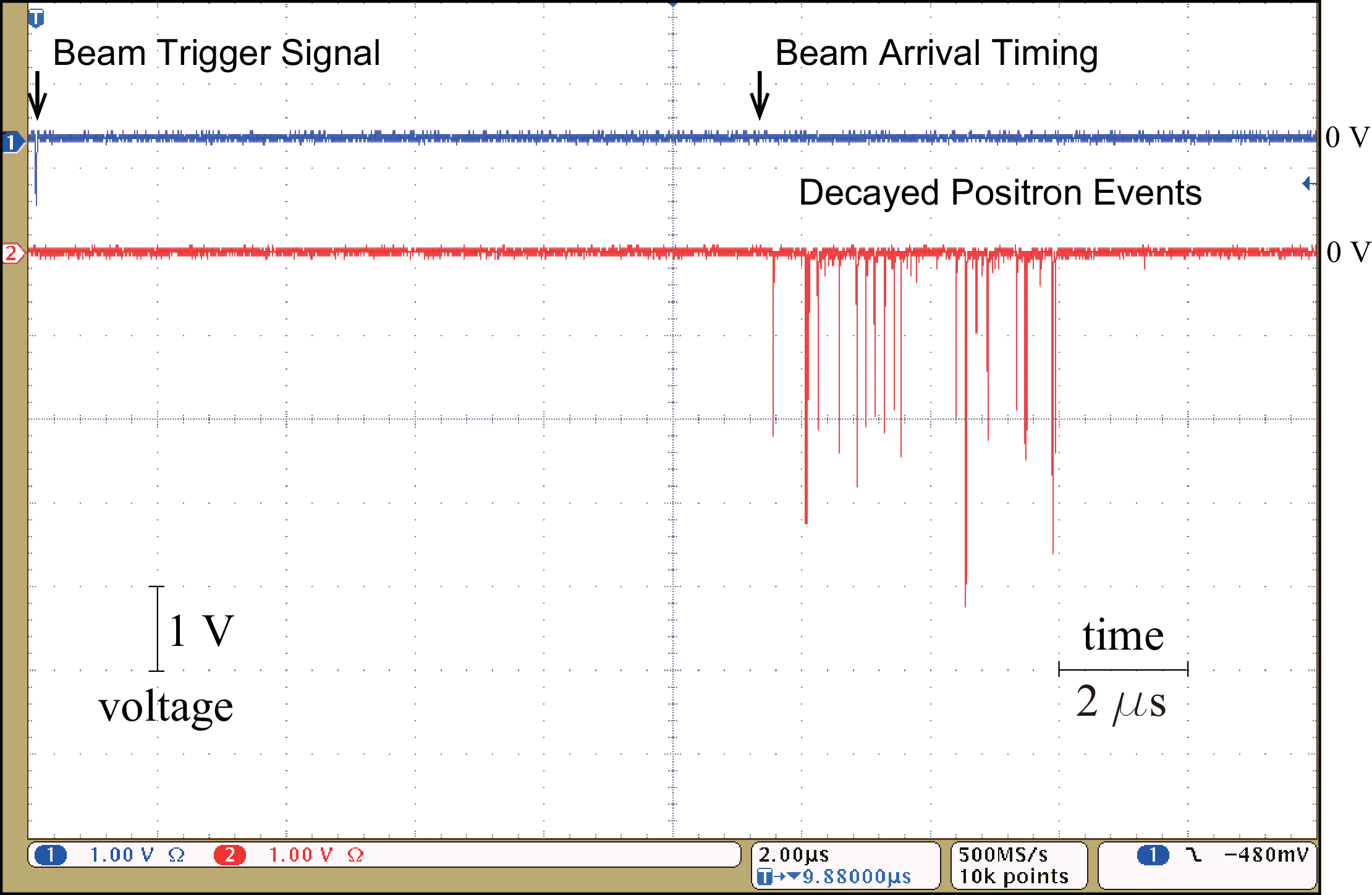} 
 \end{center}
 \caption{
Typical voltage profile of the LEFT counter at $\sim$5 Mcps with FULL-intensity beam, recorded by an oscilloscope.
The typical pulse shape is shown later in Fig.\ref{pulse-1}(a). }
 \label{scope}
\end{figure}

Fig.\ref{scope} illustrates the typical output voltage signal recorded by an oscilloscope with an input termination resistance of $R_{\rm T}=50\; \Omega$. Here, the pulse width (full width at half maximum) was $\sim 6$ ns, indicating the occurrence of pileup at periodic counting rates of over $\sim 200\;{\rm MHz}$. Notably, conventional pulse-counting techniques using voltage discriminators cannot handle $I_{\rm beam}^{\rm max}$ owing to pileup issues. In practical situations, the data-handling deadtime emerges as a bottleneck before the pileup frequency of the detector is reached. For instance, Kalliope features a deadtime of $\sim$100 ns, owing to which counting losses begin at $\sim 1/(100\;{\rm ns})\rightarrow 10$ Mcps. For random arrival times, significant counting losses begin at lower frequencies, that is, at a counting rate of $\sim$1 Mcps. 
The Kalliope system indeed experiences such a counting loss at a typical beam intensity of $10^4 \;{\rm ppp} \sim I_{\rm beam}^{\rm NORMAL}$, prompting users to set this as the maximum beam intensity.

To overcome this counting-rate bottleneck, we developed a current-readout technique for lifetime and $\mu$SR experiments. This technique is immune to pileup and deadtime effects. The remaining bottlenecks include the intrinsic saturation of the PMTs and the maximum acceptable voltage of the readout device.

\begin{figure}[htbp]
 \begin{center}
  \includegraphics[width=100mm]{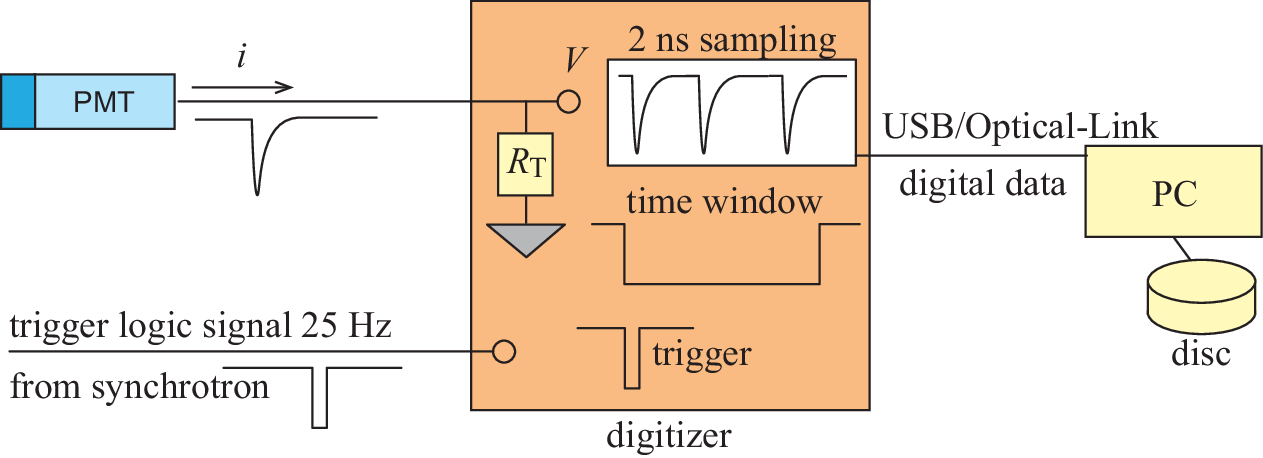}  
 \end{center}
 \caption{
Block diagram of the data taking system using a digitizer.}
 \label{block-diagram}
\end{figure}

For this purpose, we adopted a flash analog-to-digital converter (FADC) waveform digitizer (CAEN DT5370SB \cite{CAEN}) to record the output voltage waveforms of the PMTs, whose inputs were terminated using resistors with a resistance of $R_{\rm T}=50\;\Omega$. 
The adopted digitizer features a sampling rate of $f_{\rm s}=$ 500 MHz
 (sampling time step is $\Delta t=2$ ns)
, along with a 14-bit voltage resolution and eight input channels. Recorded data stored within the buffer memory (5.12 M samples/channel) of the digitizer can be transferred to computers via a USB/Optical-Link.

Considering the data transfer speeds of Optical-Link (80 Mbps) or USB 2.0 (30 Mbps), we set the data-acceptance time window to  $T_{\rm w}= 150 \,\upmu{\rm s}$, synchronizing it with the data recording trigger signal instant ($t=0$). The trigger signals were generated based on the beam-pulse timing at $f_{\rm cycle}$, provided by the 3 GeV synchrotron control signal.
The value of $T_{\rm w}$ was chosen to adequately observe the desired scientific phenomena and allow data transfer to a storage device at the required speed.
The digitized data are recorded as a time-sequential voltage data $V_j (t=k\Delta t), \, (k=0,1,2,...)$ for each $j$-th trigger signal during a time window $0\leq t \leq T_{\rm w}$.

%%%%%%%%%%%%%%%%%%%%%%%%%%%%%%%%%%%%%%%%%%%%%%%%%%%%%%%%%%%%%%%%%%%%%%%%%%%%%%%%%%%%%%

%%%%%%%%%%%%%%%%%%%%%%%%%%%%%%%%%%%%%%%%%%%%%%%%%%%%%%%%%%%%%%%%%%%%%%%%%%%%%%%%%%%%%%
%%%%%%%%%%%%%%%%%%%%%%%%%%%%%%%%%%%%%%%%%%%%%%%%%%%%%%%%%%%%%%%%%%%%%%%%%%%%%%%%%%%%%%

\section{Principle}
\label{Principle}
If we are interested in investigating relatively slow phenomena with a timescale on the order of  $\upmu$s, as in the current study, a time resolution of  $\sigma_t\sim100 \;{\rm ns}$, for instance, is sufficient to observe the time spectra of  $n(t)$. In such cases, identifying discrete pulse signals with ns precision, as required in conventional event-by-event pulse counting, is unnecessary. Instead, obtaining their averaged flux (event rate) information (i.e., current) as a function of time $t$ with a resolution of $\sigma_t$ is adequate. Therefore, in this study, we attempted to obtain flux information from the output electric currents  $i(t)$ of the PMTs.

%The number of photons corresponding to a single positron penetration event typically fluctuates owing to the changing path lengths of positrons penetrating the scintillators, statistical variations in the number of photons, and statistical fluctuations in the number of secondary electrons generated within the PMTs. These complex pulse-to-pulse charge fluctuations can be avoided through discrimination when applying the conventional pulse-counting method. On the other hand, these analog fluctuations are anticipated to influence the final precision of the flux measurement in the present method. Therefore, the final precision for $n(t)$ may be worse than that in the conventional pulse-counting method; consequently, we must quantitatively estimate the contributions of these additional fluctuations.
The number of photons produced during a single positron penetration event typically varies due to several factors, including the changing path lengths of positrons penetrating the scintillators, statistical variations in the number of photons, and statistical fluctuations in the number of secondary electrons generated within the PMTs. While these complex pulse-to-pulse charge fluctuations are mitigated through discrimination in the conventional pulse-counting method, they are anticipated to influence the final precision of the flux measurement when using the current method. Therefore, the final precision for $n(t)$ in this approach may be worse than that achieved using the conventional pulse-counting method. As a result, we must quantitatively assess the impact of these additional fluctuations on the precision of the flux measurement.

Let us begin by examining the principle.
%The basic idea of this current-readout method is that if we sum up the time-sequential voltage data $V_j(t)$ at $j$-th beam-pulse timing over multiple beam-pulses $j=1,..., N_{\rm pulse}$, then this $V_{\rm sum}(t) =-\Sigma_j V_j (t)$ must give us a spectrum proportional to the expected decay curve $n(t)$, which we usually obtain as a counting histogram $N(t)$ using the conventional pulse-counting method. Here, $N_{\rm pulse}$ denotes pulse numbers contained in a data-taking run.
The basic idea of this current-readout method is that if we sum up the time-sequential voltage data, $V_j(t)$, recorded at the $j$-th beam-pulse timing over multiple beam pulses ($j = 1, \dots, N_{\rm pulse}$), then this $V_{\rm sum}(t) = -\Sigma_j V_j (t)$ will yield a spectrum proportional to the expected decay curve $n(t)$, which is typically obtained as a counting histogram, $N(t)$, using the conventional pulse-counting method. Here, $N_{\rm pulse}$ represents the number of pulses contained in a data-taking run

%Single hit count generates a unit output charge $Q_0$ at the PMT. Ideally, it should be a constant, but it must contain fluctuation $\sigma_{Q_0}$ and noise in realistic cases. The relative fluctuation is expressed as $\hat{\sigma}_{Q_0}=\sigma_{Q_0}/Q_0$ by labeling a hat (same notations are used in the following expressing relative uncertainties). A counting rate $n$ corresponds to a current $i=dQ/dt=n Q_0$. This current is read as a voltage $V=i R_T=n R_T Q_0 $ between the termination resister $R_T$. As a result, the time spectrum of $V(t)$ becomes proportional to $n(t)$ as $V(t)=R_T Q_0 n(t)$.
A single hit count produces a unit output charge $Q_0$ at the PMT. In an ideal scenario, this charge would remain constant; however, in practice, it is subject to fluctuations $\sigma_{Q_0}$ and noise. The relative fluctuation is expressed as $\hat{\sigma}_{Q_0} = \sigma_{Q_0} / Q_0$, where the hat denotes relative uncertainties (a notation used consistently throughout the text). A counting rate $n$ corresponds to a current $i = dQ/dt = n Q_0$. This current is converted into a voltage $V = i R_{\rm T} = n R_{\rm T} Q_0$ across the termination resistor $R_{\rm T}$. Consequently, the time spectrum of $V(t)$ becomes proportional to $n(t)$, as expressed by $V(t) = R_{\rm T} Q_0 n(t)$.

%Such voltage time-spectrum must be expressed as a histogram with a binning time step of $\Delta T$. $\Delta T$ does not have to be the same as the sampling time step of the digitizer $\Delta t=2$ ns. Here, we suppose a $\Delta T$, which is wide enough to contain the width of the signal pulse. If one event exists during the $\Delta T$ period, this time bin's mean voltage is $\overline{V}_0 = R_T Q_0/\Delta T$. As for event rate spectrum, $\overline{n}_0 = 1/\Delta T$ is the mean counting rate during the $\Delta T$ period. 
The voltage time-spectrum must be expressed as a histogram with a binning time step of $\Delta T$. Importantly, $\Delta T$ does not need to match the digitizer's sampling time step $\Delta t = 2$ ns. Instead, $\Delta T$ should be wide enough to accommodate the width of the signal pulse. If a single event occurs during the $\Delta T$ period, the mean voltage in this time bin is given by $\overline{V}_0 = R_{\rm T} Q_0 / \Delta T$. Similarly, for the event rate spectrum, the mean counting rate during the $\Delta T$ period is $\overline{n}_0 = 1 / \Delta T$.

%These unit $\overline{V}_0$ and unit $\overline{n}_0$ are going to be summing up for $j=1,...,N_{\rm pulse}$ beam pulses at each $t$ including $N(t)$ events in total, yielding $\overline{V}_{\rm sum}(t)=\overline{V}_0 N(t)$ and $\overline{n}_{\rm sum}(t)=\overline{n}_0 N(t)$.
These unit values, $\overline{V}_0$ and $\overline{n}_0$, are summed over $j = 1, \dots, N{\rm pulse}$ beam pulses at each $t$, which include $N(t)$ events in total. This yields $\overline{V}_{\rm sum}(t)=\overline{V}_0 N(t)$ and $\overline{n}_{\rm sum}(t)=\overline{n}_0 N(t)$.
%The total count contained in the time bin $\Delta T$ at $t$ after the $j$-summation obtained as $N(t)=\overline{n}_{\rm sum}(t) \Delta T$ corresponds to conventional counting-histograms using TDCs.
The total count within the time bin $\Delta T$ at time $t$ after summing over $j$ is obtained as $N(t)=\overline{n}_{\rm sum}(t) \Delta T$, which corresponds to conventional counting histograms generated using TDCs.
Here we see
\begin{equation}
\overline{V}_{\rm sum}(t)=\beta \, N(t)
\label{def-beta}
\end{equation}
with $\beta=\overline{V}_0$. It means that voltage sum $\overline{V}_{\rm sum}(t)$ is proportional to the conventional counting histogram $N(t)$.

As for the fluctuations, we see 
\begin{eqnarray}
\overline{V}_{\rm sum}&=&(\beta \pm \sigma_\beta\pm \delta_{\beta})+(\beta \pm \sigma_\beta\pm \delta_{\beta})+... ({\rm for} \,N\,{\rm counts}) \nonumber \\
&=& (N\pm \sigma_N) \beta  \pm \sqrt{N\pm \sigma_N} \sigma_\beta \pm (N\pm \sigma_N) \delta_{\beta} \nonumber \\
&\sim& (N\pm \sigma_N) \beta  \pm \sqrt{N} \sigma_\beta \pm N \delta_{\beta} \nonumber \\
&=& N\beta \pm \sqrt{ \sigma_N^2 \beta^2 + N \sigma_\beta^2  + N^2 \delta_{\beta}^2  }\;\;, \\
\label{Vsum-error}
\nonumber
\end{eqnarray}
%here, $\sigma_N=\sqrt{N}$ is the Poisson statistical error, $\sigma_\beta$ is the random fluctuation of pulse height, and $\delta_{\beta}$ is a systematic error causing the voltage deviation at the same timing $t$. Error propagation of the random errors is treated as their root sum square; on the other hand, combined errors of systematic errors are treated as their simple summation. This is because systematic errors should have a common contribution without being smeared.
where $\sigma_N = \sqrt{N}$ represents the Poisson statistical error, $\sigma_\beta$ denotes the random fluctuation of pulse height, and $\delta_{\beta}$ is a systematic error causing voltage deviations at the same timing $t$. The random errors are propagated as their root sum square, whereas systematic errors are combined by simple summation. This distinction arises because systematic errors have a common contribution and do not average out.

The combined uncertainty, including the Poisson statistical fluctuation, the random fluctuation, and the systematic error, is
\begin{eqnarray}
\sigma_{\overline{V}_{\rm sum}}&=& \sqrt{ N \beta^2 (1 + \hat{\sigma}_\beta^2)  + N^2 \delta_{\beta}^2 } \nonumber \\
&=& \sqrt{\overline{V}_{\rm sum}} \sqrt{ \beta (1 + \hat{\sigma}_\beta^2 )  +  \overline{V}_{\rm sum} \hat{\delta}_{\beta}^2 }  
\label{eq-err} \\
&\rightarrow& \alpha \, \sqrt{\overline{V}_{\rm sum}}\;\;\; (\hat{\delta}_{\beta} \rightarrow 0), 
\label{eq-sqrt} \\
\nonumber
\end{eqnarray}
with
\begin{equation}
\alpha=\sqrt{ \beta (1 + \hat{\sigma}_\beta^2 )} = \sqrt{ \frac{R_{\rm T} Q_0}{\Delta T} (1+\hat{\sigma}_{Q_0}^2)}.
\label{alpha-def}
\end{equation}
%Now, we obtained an important prediction: the voltage sum's fluctuation is proportional to its square root if the systematic error is negligible. By determining the factor $\alpha$, we can estimate the random error bars of the voltage spectrum.
This result yields an important prediction: the fluctuation in the voltage sum is proportional to its square root if the systematic error is negligible. By determining the factor $\alpha$, we can estimate the random error bars of the voltage spectrum.

The following relation between relative uncertainties,
\begin{equation}
\hat{\sigma}_{\overline{V}_{\rm sum}}^2=\frac{1+ \hat{\sigma}_{\beta}^2}{N}  + \hat{\delta}_{\beta}^2,
\label{eq-relerr}
\end{equation}
obtained using Eqs.(\ref{def-beta}) and (\ref{eq-err}) 
helps us understand the expected precision in the current-readout method, including the systematic error.
The conventional pulse counting method can be understood as a particular case of Eq.(\ref{eq-relerr}) with $\hat{\sigma}_{\beta}=0$ and $\hat{\delta}_{s}=0$, resulting a well-known relation of $\hat{\sigma}_N=1/\sqrt{N}$.

The random error due to random fluctuation of the generated photon number inside the scintillator, the length of the path length, gain fluctuation, and random electric noise contribute to $\hat{\sigma}_{\beta}$. 
%However, it affects only increasing the factor $\alpha$. We can still expect to improve the precision by increasing the statistics $N$.
However, these factors only increase the value of $\alpha$, and the precision can still be improved by increasing the statistics $N$.

%An example of the systematic error $\hat{\delta}_{\beta}$ is a dumping oscillation of the ground-level voltage, synchronizing with the trigger timing $t=0$. The reproducibility of this oscillation does not cancel each other but systematically produces voltage deviation on $\overline{V}_{\rm sum}$ from zero. Eq.(\ref{eq-relerr}) implies that the precision will be limited due to this systematic error if $\hat{\delta}_{\beta}$ is not negligible. In that case, the square root property of Eq.(\ref{eq-sqrt}) will be violated, and $\sigma_{\overline{V}_{\rm sum}}$ will show a more substantial increase as a function of $\overline{V}_{\rm sum}$.
An example of a systematic error $\hat{\delta}_{\beta}$ is the damped oscillation in the ground-level voltage, synchronized with the trigger timing at $t=0$. Since this oscillation is reproducible, it does not cancel out and systematically causes deviations in $\overline{V}_{\rm sum}$ from zero. Eq. (\ref{eq-relerr}) suggests that the precision will be limited by this systematic error if $\hat{\delta}_{\beta}$ is significant. In such cases, the square root behavior predicted by Eq. (\ref{eq-sqrt}) will no longer hold, and $\sigma_{\overline{V}{\rm sum}}$ will increase more substantially as a function of $\overline{V}_{\rm sum}$.

Experimentally, the quantities used in this section are obtained as follows.
The time-sequential voltage data $V_{j}(t)$ is obtained at the digitizer for $t=k\Delta t$ ($k=0,1,2,...,T_{\rm w}/\Delta t$-1), starting at the $j$-th beam-pulse arrival timing of $t=0$.
Then, using
\begin{equation}
V_{\rm sum}(t)=-\sum_{j=1}^{N_{\rm pulse}} V_j(t), \;\; {\rm for}\;\; t=k\Delta t \,,
\end{equation}
$\overline{V}_{\rm sum}$ is obtained by rebinning (combining bins) $m$-bins and dividing it by $m$ as
\begin{equation}
\overline{V}_{\rm sum}(t=K\Delta T)=\sum_{k'=0}^{m-1} \frac{V_{\rm sum}(K\Delta T+k'\Delta t)}{m}
, \;\; {\rm with} \;\; K=0,1,2,...,T_{\rm w}/\Delta T-1 \,
\end{equation}
to obtain the mean summed voltage during the rebinning time $\Delta T$.
In the following analysis, $\Delta t=$2 ns, $\Delta t_{\rm rebin}=\Delta T=m \Delta t=20$ ns, and $m=10$ will be used.

%Note that $\alpha$ defined in Eq.(\ref{alpha-def}) is not a dimensionless parameter but contains $\sqrt{\rm V}$, where $\rm V$ is volt (unit). Considering that $\hat{\sigma}_\beta$ is unknown, directly estimating $\alpha$ based on the fluctuations observed in the experimental data of $\overline{V}_{\rm sum}$ appears to be a better alternative compared to using Eq.(\ref{alpha-def}).  Also, $\alpha_{\rm rebin}(m)=\alpha (m=1)/\sqrt{m}$ will be used using Eq.(\ref{alpha-def}). Further discussions on this aspect will be outlined in Section \ref{Discussion}.
It is important to note that $\alpha$, as defined in Eq. (\ref{alpha-def}), is not dimensionless and includes a unit of $\sqrt{\text{V}}$, where V represents volts. Given that $\hat{\sigma}_{\beta}$ is typically unknown, estimating $\alpha$ directly from the observed fluctuations in the experimental data for $\overline{V}_{\rm sum}$ is a more practical alternative than relying on Eq. (\ref{alpha-def}). Moreover, $\alpha_{\rm rebin}(m) = \alpha (m=1)/\sqrt{m}$ will be used, based on Eq. (\ref{alpha-def}). Additional discussions on this approach are provided in Section \ref{Discussion}.

%%%%%%%%%%%%%%%%%%%%%%%%%%%%%%%%%%%%%%%%%%%%%%%%%%%%%%%%%%%%%%%%%%%%%%%%%%%%%%%%%%%%%%

\section{Simulation}
\label{Simulation}
A Monte-Carlo simulation study was conducted to illustrate the behavior of $V_{\rm sum}(t)$. 
Signal pulses with an asymmetric Gaussian shape were generated, characterized by a rise time of 2 ns and a fall time of 4 ns. This pulse shape closely resembles that of real signals. 
The mean peak voltage was set to 2 V, with $\pm 0.2$  V Gaussian fluctuations. Following this, event arrival timings were generated randomly at constant count rates of approximately (a) 10 Mcps, (b) 100 Mcps, (c) 1 Gcps, and (d) 10 Gcps. 
The corresponding pulse signals were recorded using a sampling time step of $\Delta t = 2 \; {\rm ns}$.

\begin{figure}[htbp]
 \begin{center}
  \includegraphics[width=86mm]{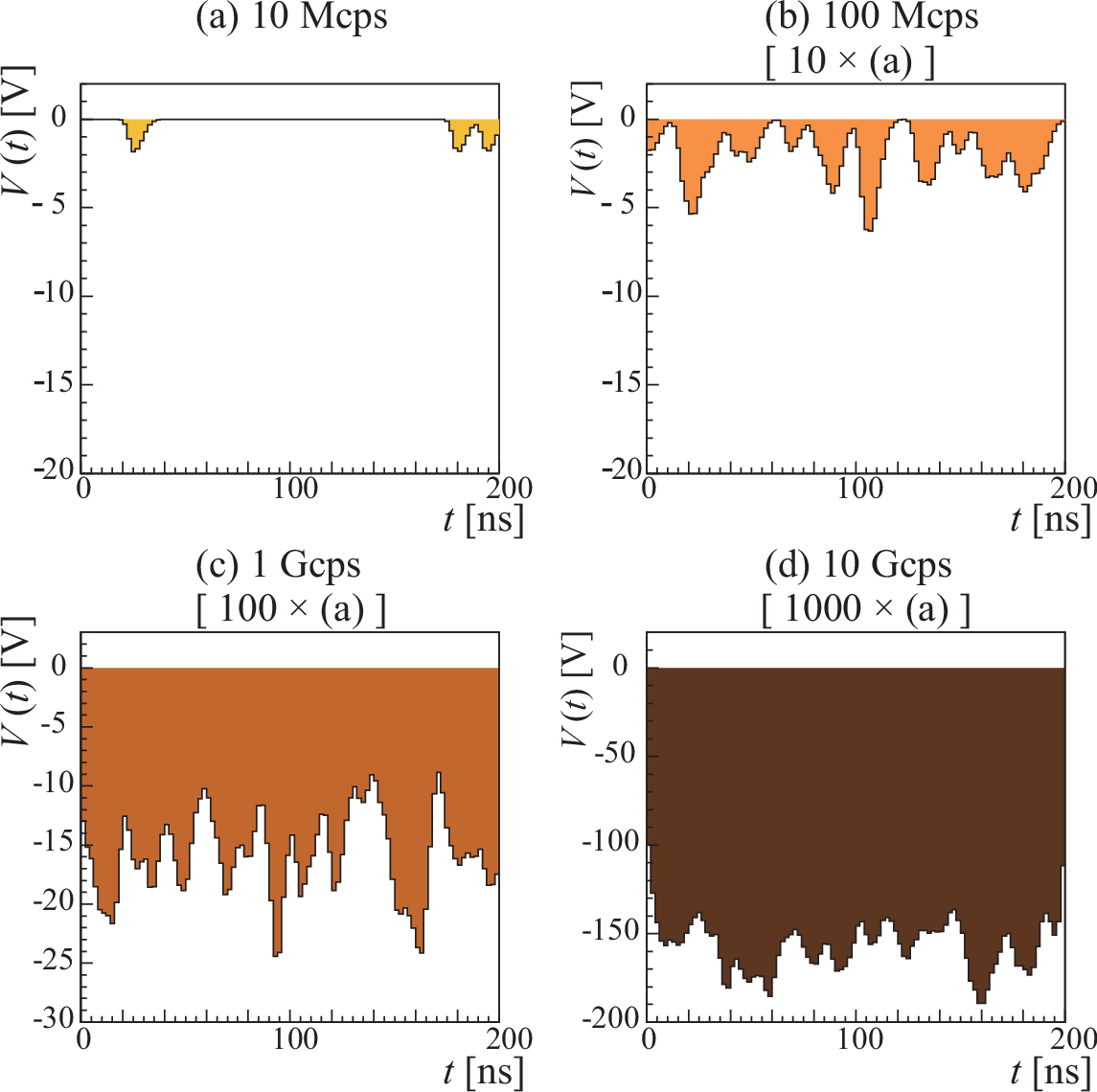} 
 \end{center}
 \caption{
 Simulation results of time-sequential voltage data $V(t)$ at different event rates.}
 \label{time-sim}
\end{figure}

The time-sequential voltage data, $V(t)$, are shown in Fig.\ref{time-sim}(a-d). These data correspond to the input voltage signals recorded by digitizers or oscilloscopes, similar to that displayed in Fig.\ref{scope}. The signal pileup phenomenon may initiate from $\sim$10 Mcps, making individual pulses with count rates of over 1 Gcps indistinguishable.

These results can also be interpreted as (a) $V(t)$ for a beam pulse at 10 Mcps or (b/c/d) $V_{\rm sum}(t)$ obtained by summing over $N_{\rm pulse}=$10/100/1000 beam pulses at 10 MHz as follows:
\begin{equation}
    V(t)^{(b/c/d)} \leftrightarrow \sum_{j=1}^{N_{\rm pulse}=10/100/1000} V(t)_{j}^{(a)}.
\end{equation}
%This demonstrates the behavior of $V_{\rm sum}$.
This demonstrates how $V_{\rm sum}(t)$ arises from the summation of individual pulses $V_j(t)$.

%Evidently, voltage discrimination is impossible for count rates exceeding 1 Gcps owing to severe pileup phenomena. However, even before reaching this threshold, pulse identification efficiency decreases owing to counting losses resulting from two-pulse pileups. In addition to this, counting losses originate from the intrinsic deadtime of PMTs resulting from effective bias reduction and the electric deadtime attributed to the processing of downstream digital signals in the conventional pulse-counting method utilizing discriminators. Consequently, counting losses initiate even before reaching the count rate at which signal pileup begins, which can be estimated based on the signal width.
For count rates above 1 Gcps, voltage discrimination becomes infeasible due to severe pileup effects. Additionally, even before reaching this limit, counting efficiency decreases due to two-pulse pileup losses. Furthermore, intrinsic PMT deadtime (resulting from bias reduction) and electronic deadtime (arising from signal processing in conventional pulse-counting methods) contribute to additional counting losses. These effects initiate at count rates lower than those causing visible pileups, as estimated from the signal width.

\begin{figure}[htbp]
 \begin{center}
  \includegraphics[width=86mm]{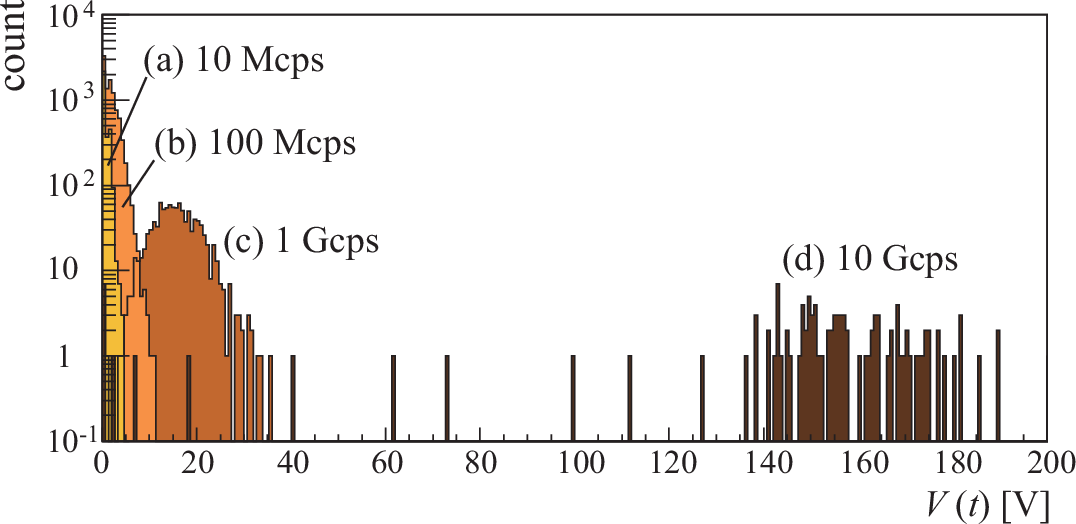} 
 \end{center}
 \caption{
Voltage distributions obtained by projecting the $V(t)$ data plotted in Fig.\ref{time-sim} along the voltage axis.}
 \label{hist-sim}
\end{figure}

To examine the voltage distributions, we obtained the time-projection distributions of the profiles shown in Fig.\ref{time-sim}, and the obtained distributions are displayed in Fig.\ref{hist-sim}. For case (a), the shown distribution corresponds to the voltage profile of a typical single pulse whose endpoint lies at a peak voltage of 2 V. In cases (b-d), the mean voltage values increase due to accidental signal pileups. The minimum voltage of the distributions exceeds zero, particularly for count rates exceeding 1 Gcps (c, d). This implies that the time-sequential signal is consistently piled up without returning to zero, i.e., $V(t)>0$ for any $t$. As shown in Fig.\ref{time-sim}(c,d), pulse-like structures remain even under high-counting-rate conditions without being smeared. The amplitudes of these ``pulses'' can be as large as 10 V. Hence, these cannot be identified as single signal pulses, known to exhibit peak heights of $\sim$2 V. This unnatural behavior will be discussed in Section \ref{Discussion}.

\begin{figure}[htbp]
 \begin{center}
  \includegraphics[width=86mm]{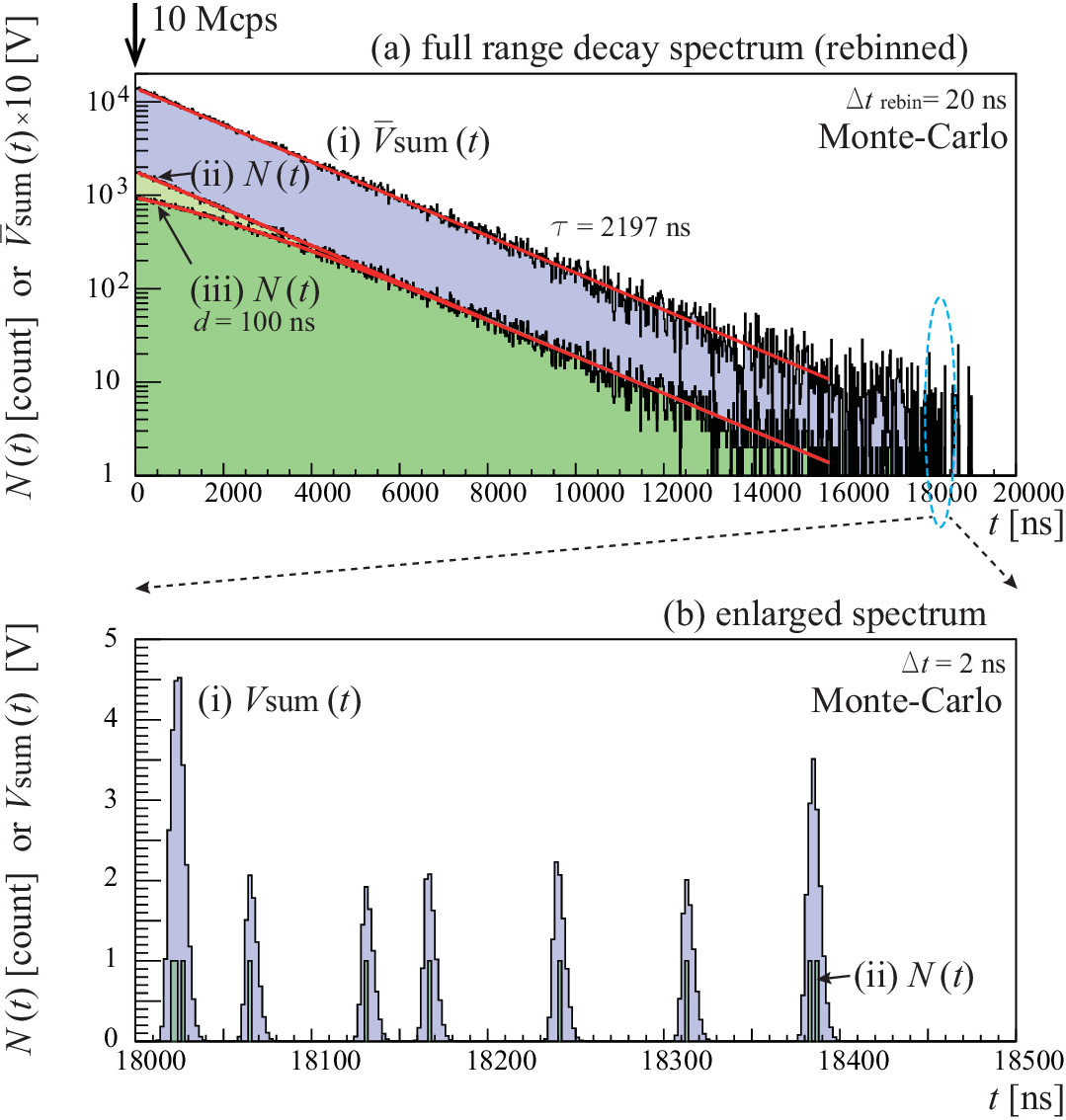} 
 \end{center}
 \caption{
%Decay curve simulation results. (a) $\overline{V}_{\rm sum}(t) \times 10$ [V], $N(t)$ [count], and $N(t)$ [count] with dead time effect for $d=100$ ns are plotted with a time bin of $\Delta t_{\rm rebin}$ $N(t)$ is also rebinned for this plot. (b) $N(t)$, $V_{\rm sum}(t)$ is plotted instead of $\overline{V}_{\rm sum}(t)$ with $\Delta t$ without rebining.
Decay curve simulation results. (a) The quantities $\overline{V}_{\rm sum}(t) \times 10$ [V], $N(t)$ [count], and $N(t)$ [count] with the dead-time effect for $d = 100$ ns are plotted using a time bin of $\Delta t_{\rm rebin}$. For comparison, $N(t)$ is also rebinned for this plot.
(b) The values of $N(t)$ and $V_{\rm sum}(t)$ are plotted instead of $\overline{V}_{\rm sum}(t)$, using $\Delta t$ without rebinning.
}
 \label{decay-sim}
\end{figure}

To demonstrate this current-readout method, we also performed a simulation study generating events that obey a frequency proportional to exponential decay at parameters similar to real measurements.
The rate function was set as
$
n(t)=n_0 e^{-t/\tau_\mu},
$
by setting $n_0=10$ Mcps and $\tau_\mu=2197$ ns.
Voltage data $V(t)$ were generated as same as for Fig.\ref{time-sim}, for $j=1,...,10000$.
Obtained results are shown in Fig.\ref{decay-sim} (a).
$\overline{V}_{\rm sum}(t)$ and corresponding counting histogram $N(t)$ are shown.
We can see their excellent agreement with each other, confirming the proportionality of Eq.(\ref{def-beta}).
As shown in Fig.\ref{decay-sim}(b), we can confirm that ${V}_{\rm sum}(t)$ represents $N(t)$ for the individual pulses.
We also simulated a deadtime effect on $N(t)$ by introducing a VETO time $d=100$ ns after each accepted count, shown in the same figure.
As expected, the decay curve reveals a saturation effect as shown in Fig.\ref{decay-sim}(b)-(iii), which shape is well reproduced with a function between accepted rate $n_{\rm accept}$ and requested rate $n_{\rm request}$ as
\begin{equation}
n_{\rm accept}=\frac{1}{1/n_{\rm request}+d}.
\label{n-deadtime}
\end{equation}
There should be no saturation on $\overline{V}_{\rm sum}$. 
Now, we confirm the expectation discussed in Section \ref{Principle}.
We will see the results of the experimental study in the following sections.

%%%%%%%%%%%%%%%%%%%%%%%%%%%%%%%%%%%%%%%%%%%%%%%%%%%%%%%%
%%%%%%%%%%%%%%%%%%%%%%%%%%%%%%%%%%%%%%%%%%%%%%%%%%%%%%%%%%%%%%%%%%%%%%%%%%%%%%%%%%%%%%

\section{Results}
\label{results}
\subsection{Charge Spectra}
\label{results-qdc}

\begin{figure}[htbp]
 \begin{center}
  \includegraphics[width=86mm]{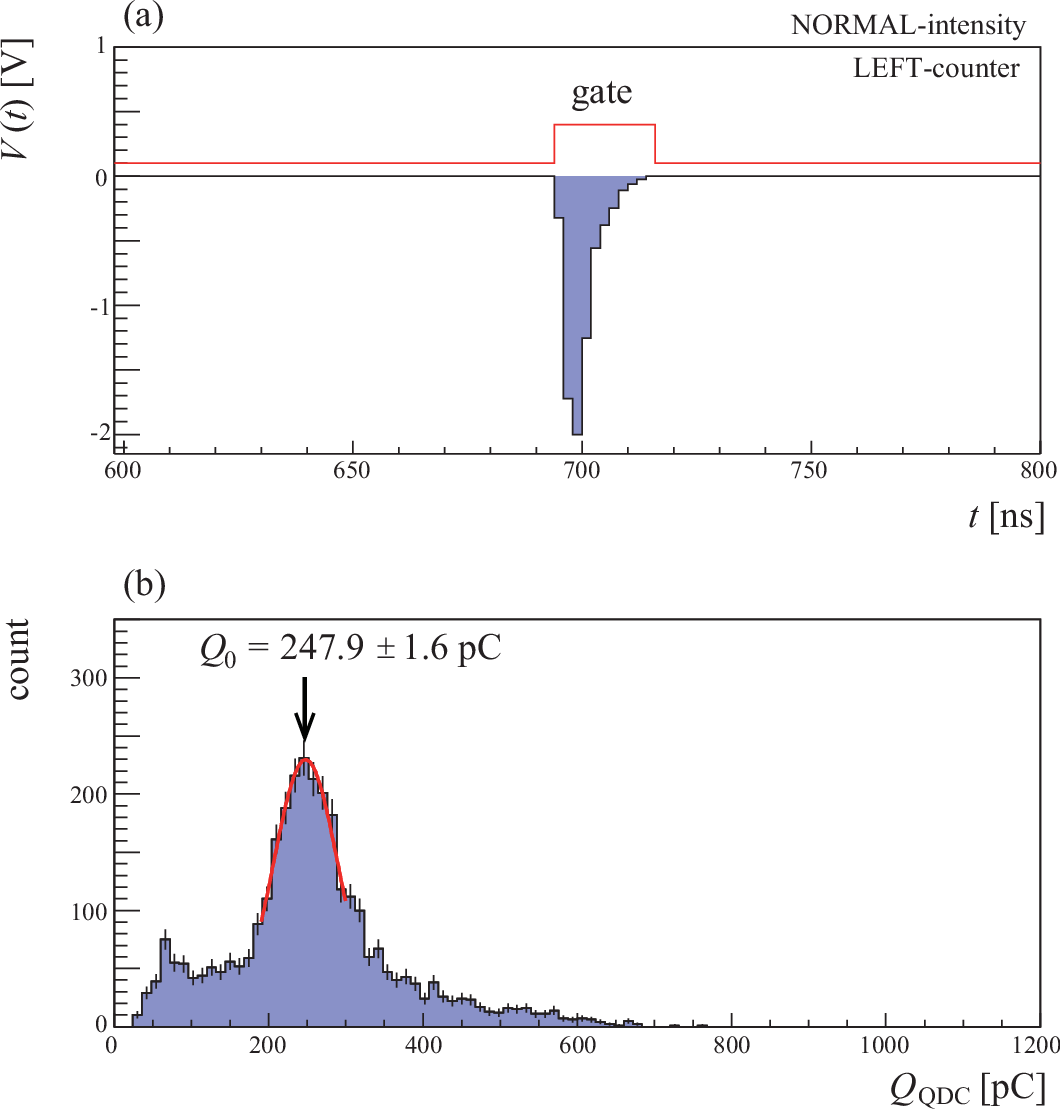}
 \end{center}
 \caption{
Data recorded by the LEFT counter under PMT exposure to the NORMAL-intensity beam. (a) Typical $V(t)$ of a PMT, recorded by a digitizer. (b) $Q_{\rm QDC}$ distribution from the ``QDC'' analysis.}
 \label{pulse-1}
\end{figure}

Fig.\ref{pulse-1}(a) shows a typical output signal extracted from a PMT detecting positrons generated from the decay of stopped muons in the $\mu$LV-Run0 experiment. 
It shows the voltage pulse shape, similar to that displayed on an oscilloscope (Figs.\ref{scope} and \ref{time-sim}). A software time gate was employed in offline data analysis to perform pulse-charge measurements similar to using QDCs, as shown in Fig.\ref{pulse-1}(b). 

Similar to hardware discrimination, software gates were determined using leading-edge discrimination. The integrated charge within the gate was calculated as follows:
\begin{equation}
Q_{\rm QDC}=-\frac{1}{R_{\rm T}} \int_{\rm gate} V(t)dt =- \frac{T_{\rm gate}}{R_{\rm T}} \sum_{ k' \; \subseteq \; {\rm gate}} V(k'\Delta t) .
\end{equation}
The distribution of this integrated charge over many events is shown in Fig.\ref{pulse-1}(b). This distribution corresponds to the charge distribution derived from conventional hardware QDC measurements. The displayed profile exhibits a peak structure corresponding to the mean energy deposition within the thin plastic scintillation counter, where almost all positrons penetrate without stopping. By determining the peak position, we can estimate the average charge output derived from the PMT and flowing through $R_{\rm T}$ as $Q_0=248 \; {\rm pC}$. Based on this, we can convert the observed integrated charge $Q$ into the equivalent hit-event number using $Q_0$ as $N_{\rm equiv.}=Q/Q_0$.

\begin{figure}[htbp]
 \begin{center}
  \includegraphics[width=86mm]{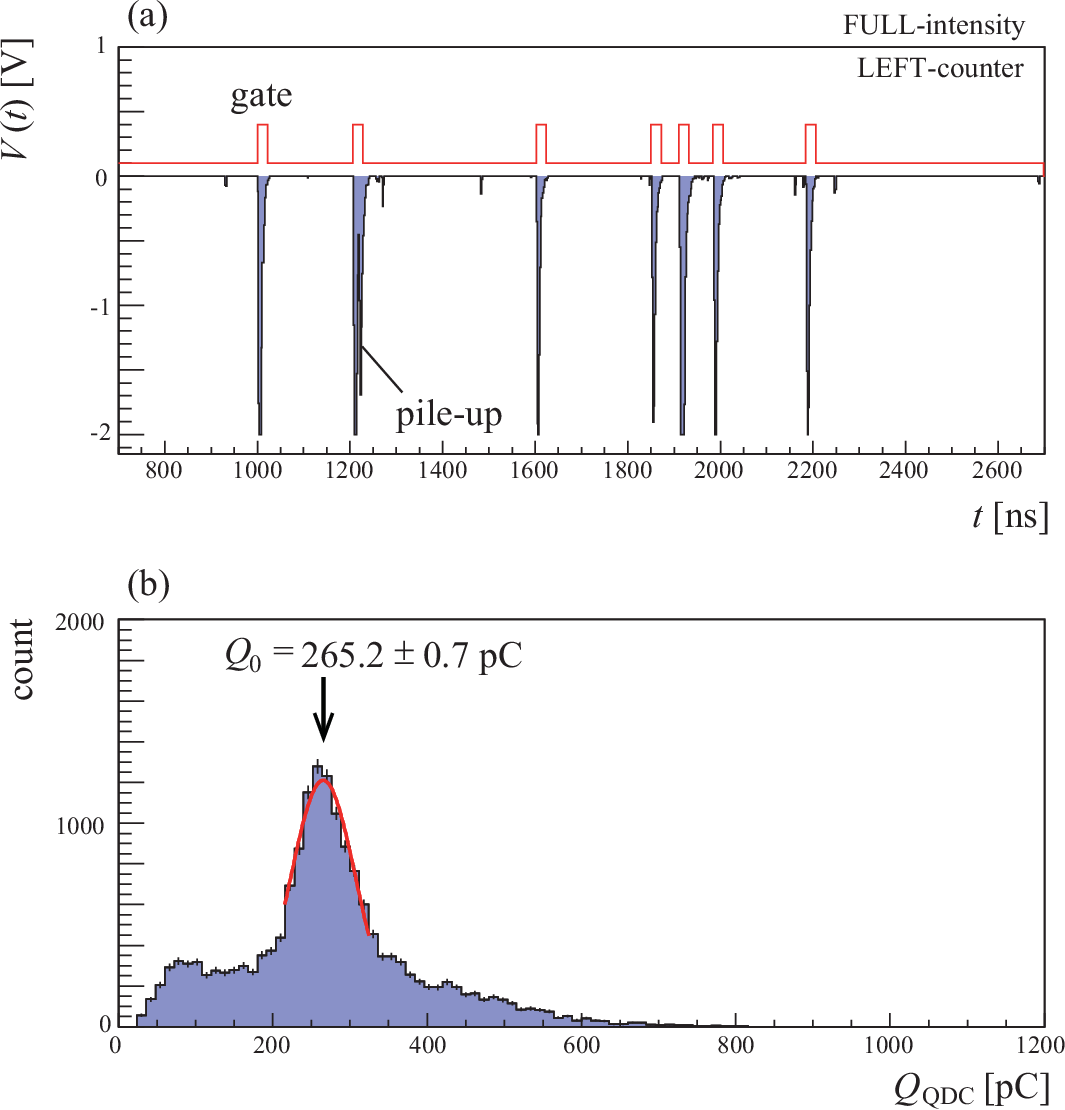}
 \end{center}
 \caption{Same as Fig.\ref{pulse-1} but for results for FULL-intensity beam in the LEFT counter are shown.}
 \label{pulse-2}
\end{figure}

\begin{figure}[htbp]
 \begin{center}
  \includegraphics[width=86mm]{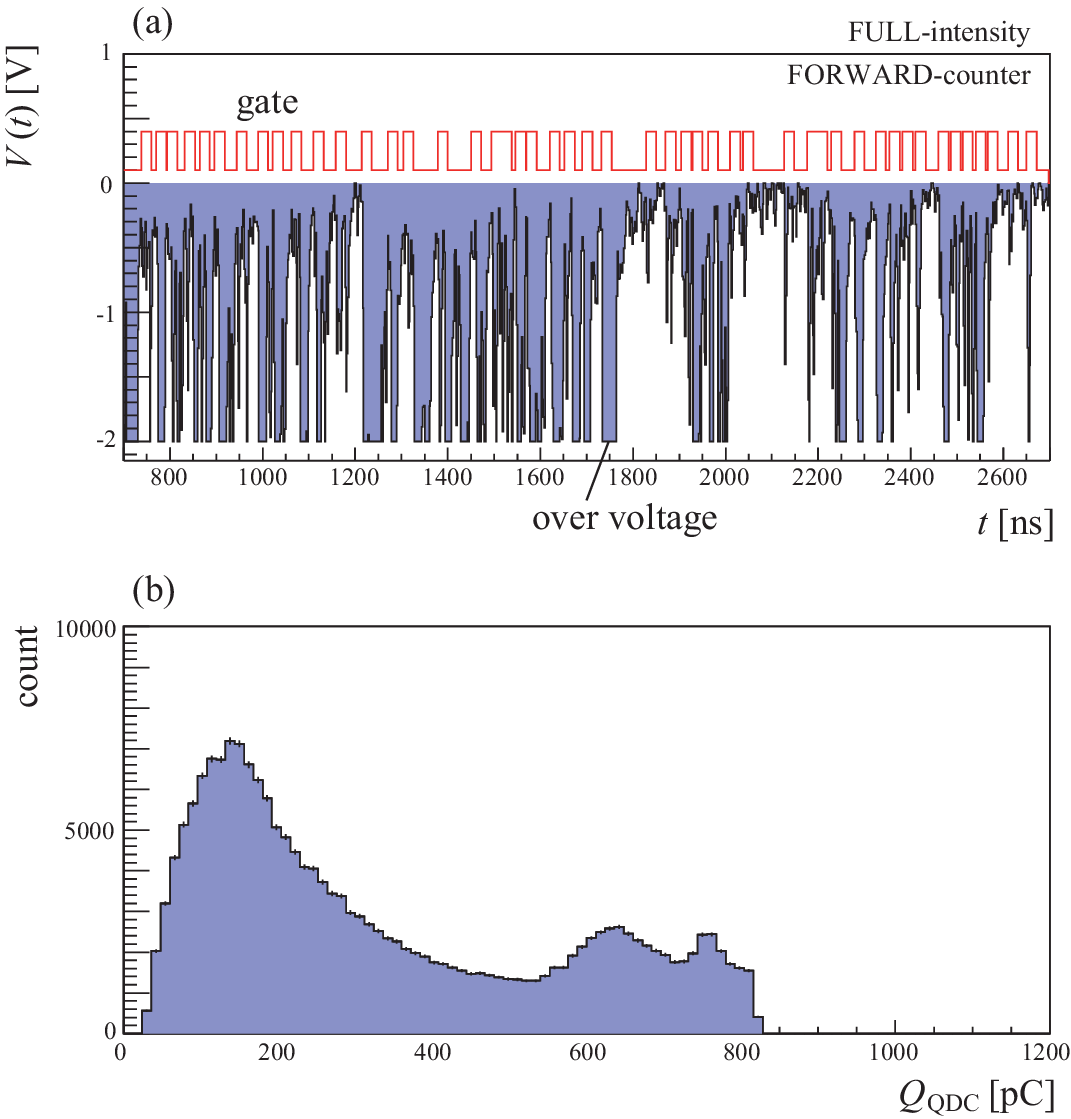}
 \end{center}
 \caption{Same as Fig.\ref{pulse-1} but for results for FULL-intensity beam in the FORWARD counter are shown. Severe pileups are apparent.}
\label{pulse-3}
\end{figure}

The data displayed in Fig.\ref{pulse-1} were recorded by the LEFT counter under PMT exposure to the NORMAL-intensity beam. Similar results were observed for the FULL-intensity beam, as shown in Fig.\ref{pulse-2}. Although signal pileups were not apparent in this case, the pulse-to-pulse duration was sufficiently small to initiate accidental pileups. The highest event rate scenario in our data was observed in the results obtained from the FORWARD counter under exposure to the FULL-intensity beam, as shown in Fig.\ref{pulse-3}. Here, we notice severe pileups, which prevent the separation of independent signal pulses. This situation roughly corresponds to the simulation result shown in Fig.\ref{time-sim}(b). Notably, the digitizer employed in the experiment has a maximum measurable voltage of 2 {\rm V}, resulting in a voltage cut-off at this level. Hence, we partially lost voltage information for well-overlapping pulses. To avoid this, the data must have ideally been recorded at $|V|\sim 1\;{\rm V}$ by adjusting the gain. Unfortunately, we were not fully aware of this voltage cut-off limitation during the $\mu$LV-Run0 experiment.

The peak value shown in Fig.\ref{pulse-2}(b) is slightly larger than that shown in Fig.\ref{pulse-1}(b), which could be due to noise increasing. Consequently, the $Q_0$ value estimated from this QDC analysis may not be adequately reliable to be treated as a constant value across different measuring conditions in subsequent data analysis.
%%%%%%%%%%%%%%%%%%%%%%%%%%%%%%%%%%%%%%%%%%%%%%%%%%%%%%%%%%%%%%%%%%%%%%%%%%%%%%%%%%%%%%
%%%%%%%%%%%%%%%%%%%%%%%%%%%%%%%%%%%%%%%%%%%%%%%%%%%%%%%%%%%%%%%%%%%%%%%%%%%%%%%%%%%%%%
\subsection{Time Spectra}
\label{results-tdc}
\begin{figure}[htbp]
 \begin{center}
  \includegraphics[width=86mm]{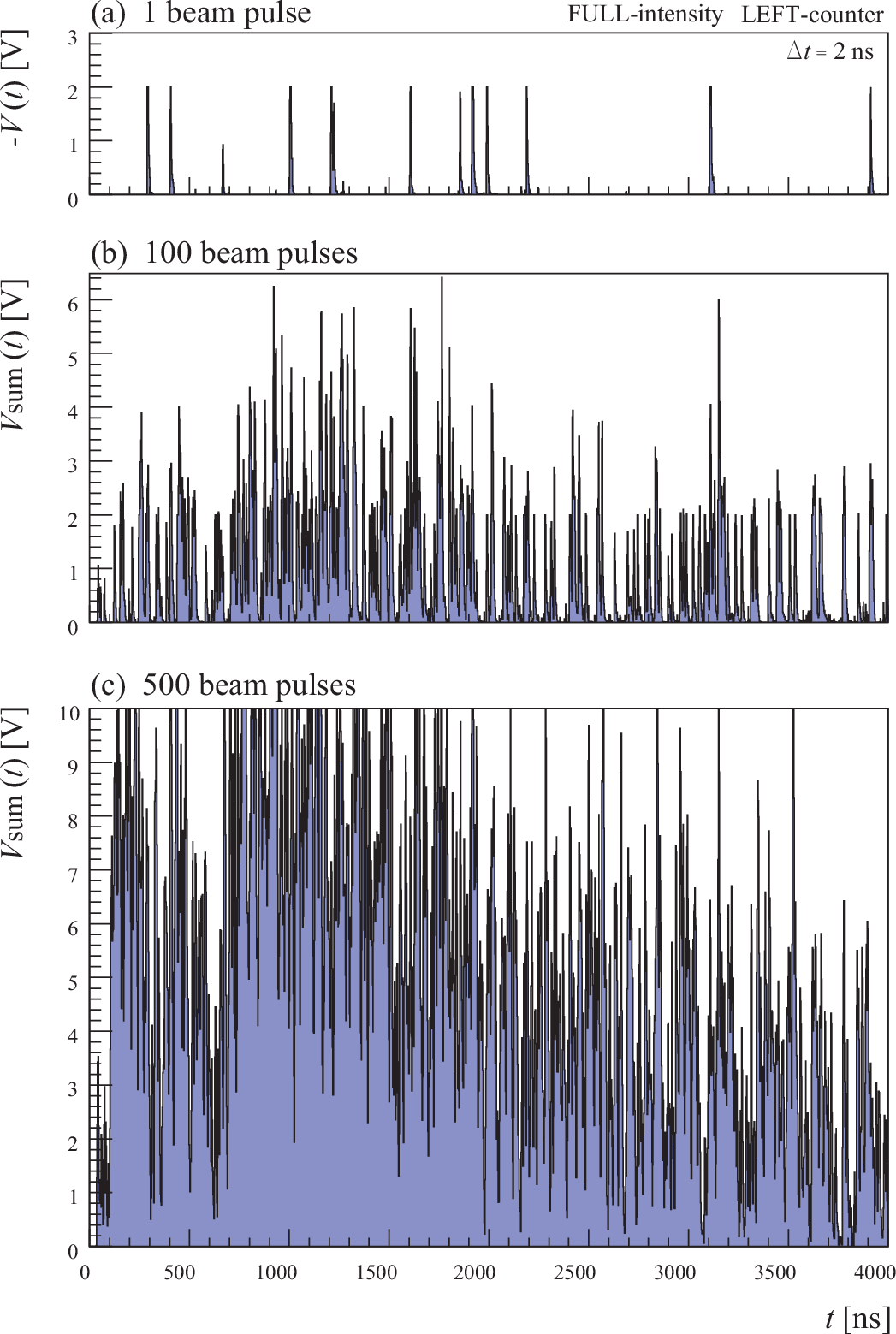}
 \end{center}
 \caption{
Data recorded by the LEFT counter under PMT exposure to the FULL-intensity beam. (a) Sample time-sequential voltage data, $-V(t)$, presenting one beam pulse. Individual event pulses are isolated. (b) $V_{\rm sum}(t)$ integrated over 100 beam pulses. ``Pileup'' (pulse overlapping) can be observed. (c) $V_{\rm sum}(t)$ integrated over 500 beam pulses. ``Pileup'' events form a ``decay curve.''}
 \label{decay-pulse}
\end{figure}

%The primary scientific measurement in this study focused on the histogram generated by the conventional pulse-counting method. 
The primary scientific measurement in this study focused on the time spectra of event probability, which corresponds to the TDC histogram $N(t)$ generated by the conventional pulse-counting method. 
Samples of the obtained integrated voltages, $V_{\rm sum}(t)$, are shown in Fig.\ref{decay-pulse}, where (a, b, c) correspond to  $N_{\rm pulse}=$(1, 100, 500), i.e., corresponding data-taking run's time-length of $T_{\rm run}=N_{\rm pulse}/T_{\rm cycle}=$(1/25 s, 4 s, 20 s). 
Evidently, compared to the pulse signals displayed in Fig.\ref{decay-pulse}(a), those shown in Fig.\ref{decay-pulse}(b, c) demonstrate increased overlapping, forming spectra similar to conventional TDC histograms that reflect an exponential decay curve.

\begin{figure}[htbp]
 \begin{center}
  \includegraphics[width=86mm]{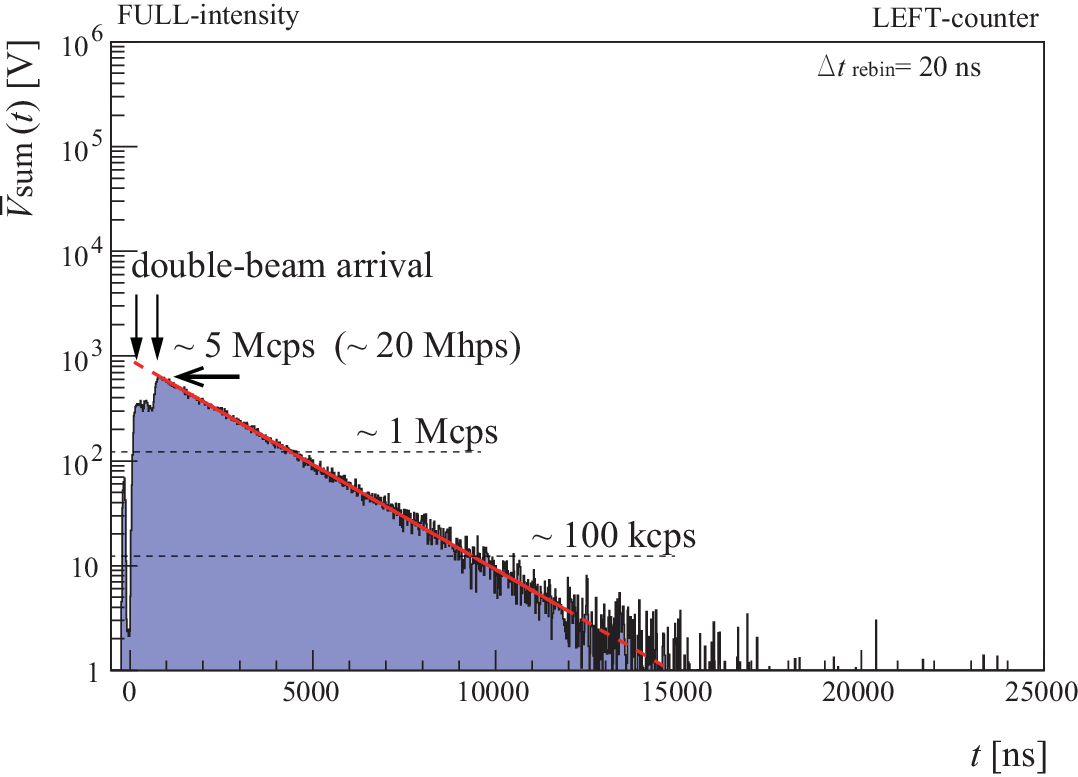}
 \end{center}
 \caption{
$\overline{V}_{\rm sum}(t)$ spectrum extracted from the LEFT counter under FULL-intensity beam exposure, representing a simple exponential decay curve. The solid line indicates the fitting result obtained using the exponential function $\overline{V}_{\rm sum}(t)=A_{\rm V} e^{-t/\tau_\mu}$. Corresponding counting rates are also included and were estimated using the $Q_0$ value. The double-pulse structure of the muon beam, with the 600-ns interval, can be seen.
The $t=0$ position is arbitrary.
%$\overline{V}_{\rm sum}(t)$ data were derived as the average values of $V_{\rm sum}(t)$ computed over a rebinning time period of $\Delta t_{\rm rebin}$=20 ns for 10 time bins.
}
 \label{decay-0G-ch1}
\end{figure}

\begin{figure}[htbp]
 \begin{center}
  \includegraphics[width=86mm]{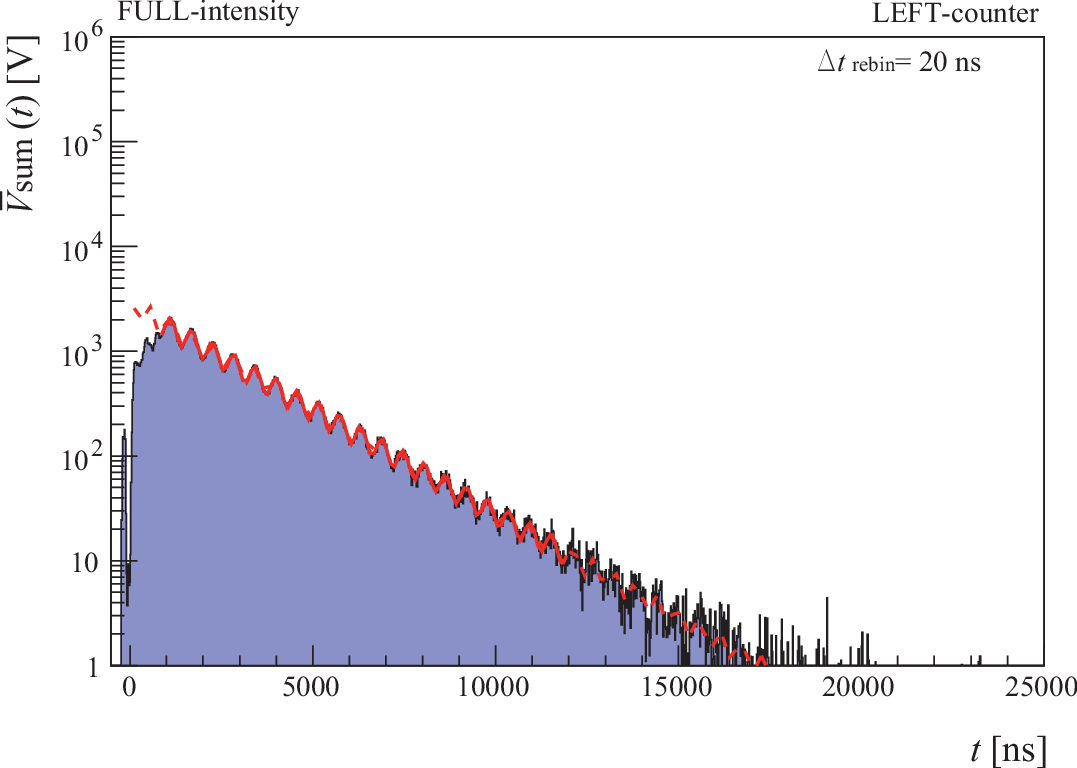}
 \end{center}
 \caption{
Same as Fig.\ref{decay-0G-ch1}, but a transverse magnetic field was applied to generate spin precession for this data. The solid line indicates the fitting result obtained using the exponential function, including spin precession
$\overline{V}_{\rm sum}(t)=A_{\rm V} [1+A_{\rm PV}\, {\rm cos} (\omega_{\rm L} t + {\rm phase})] e^{-t/\tau_\mu}$
, illustrated as the oscillation pattern around the exponential decay curve.}
 \label{decay-123G-ch1}
\end{figure}

The results of $\overline{V}_{\rm sum}(t)$ integrated over $T_{\rm run}=1$ h each are shown in Figs.\ref{decay-0G-ch1} and \ref{decay-123G-ch1}. Fig.\ref{decay-0G-ch1} displays a decay curve, which simply represents a single exponential decay events with a constant lifetime. Fig.\ref{decay-123G-ch1} presents a $\mu$SR spectrum obtained by subjecting the muon stopper to a transverse (vertical) magnetic field. By fitting the obtained data, we can obtain the lifetime and spin precession frequency. 
% for v3
The fact that the obtained integrated voltage is proportional to the scientifically expected exponential decay curve also shows the proportionality of Eq.(\ref{def-beta}).
These results indicate that the current-readout method is sufficiently capable of performing such scientific measurements. The time spectra shown here are plotted with error bars computed using Eq.(\ref{eq-sqrt}) after determining $\alpha$. The estimation of random errors, including the determination of $\alpha$, will be discussed later in Section \ref{Discussion}.

\begin{figure}[htbp]
 \begin{center}
  \includegraphics[width=86mm]{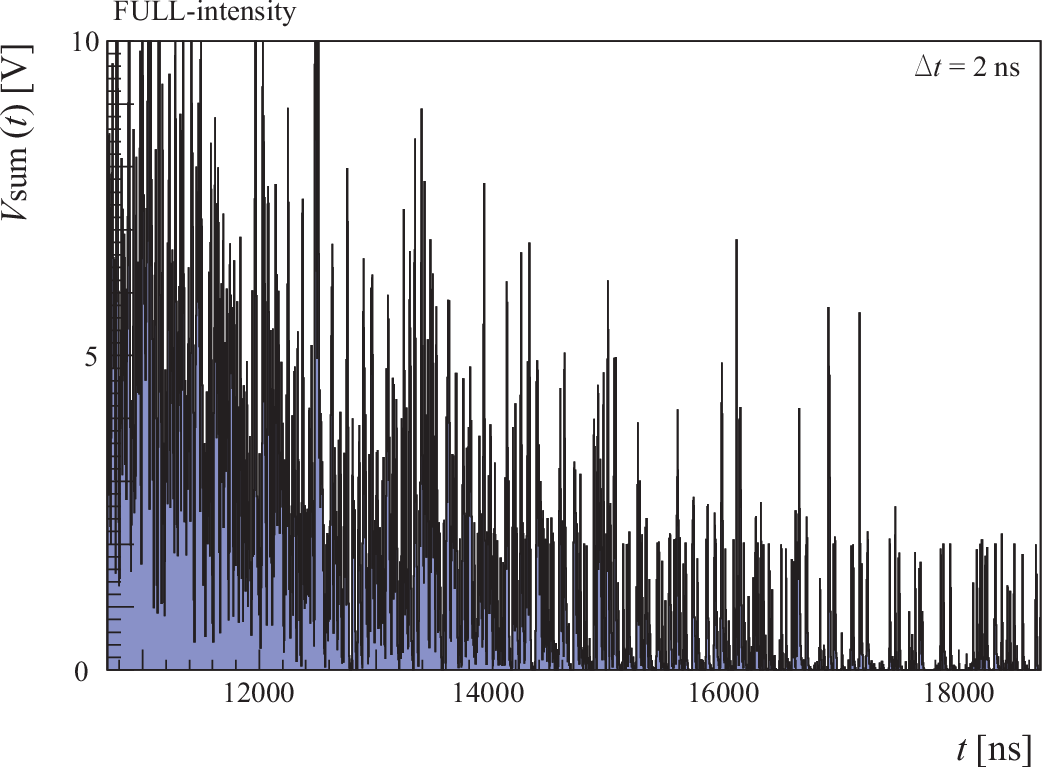}
 \end{center}
 \caption{
Same as Fig.\ref{decay-pulse} but showing a different $t$ region. Independent event pulses at $t\sim 18,000$ ns are apparent; however, their pileup begins forming a decay curve structure at $t\sim 12,000$ ns.}
 \label{decay-0G-ch1-tail}
\end{figure}

The profiles shown in both Figs.\ref{decay-0G-ch1} and \ref{decay-123G-ch1} resemble conventional TDC histograms. However, these are not histograms but integrated voltage data. %In these figures, $\overline{V}_{\rm sum}$ is plotted instead of $V_{\rm sum}$. 
%mod ver3
%$\overline{V}_{\rm sum}$ is defined as the average value of $V_{\rm sum}$ over a rebinning time period of $\Delta t_{\rm rebin}=20$ ns for 10 time bins. 
%Indeed, t
%The tail region presented in Fig.\ref{decay-0G-ch1-tail} reveals the shapes of remaining independent pulses, which do not appear in conventional TDC histograms. 
The tail region presented in Fig.\ref{decay-0G-ch1-tail} reveals the shapes of remaining independent pulses. 
%This indicates that integrated voltage spectra, $V_{\rm sum}(t)$, can be interpreted as averaged flux values $n(t)$ within the high-statistics region; conversely, discrete pulses remain unsmeared by other pulses in the low-statistics region. 
The corresponding counting-rate written in Fig.\ref{decay-0G-ch1} was estimated by comparing the voltage sum with the $Q_0$ value
 as 
\begin{equation}
n_{\rm LEFT} (t=0)=\frac{\overline{V}_{\rm sum}(t)}{\eta}, \;\;\; \eta \equiv R_{\rm T} N_{\rm pulse} Q_0.
\label{eta}
\end{equation}
Using $Q_0$, we might estimate the absolute counting number and random errors by relying on Poisson statistics. However, the measured voltages may include other random fluctuation sources beyond Poisson fluctuations. The counting rates shown in the following discussions are estimated using Eq.(\ref{eta}) to see rate stabilities, but they will not be used to estimate the statistical error. 
The subsequent section will detail a direct analysis of these random fluctuations without using the exact counting numbers.

%where $K$ denotes the number like $k$, but of the rebinned data $\overline{V}_{\rm sum}$, i.e., $t=K \Delta t_{\rm rebin}$.
%\begin{equation}
%n_{\rm LEFT} (t=0)=\frac{\sum_K^{t \subseteq {\rm decay curve}} \overline{V}_{\rm sum}(t) \Delta t_{\rm rebin}}{R_{\rm T} N_{\rm pulse} Q_0 %\tau_\mu},
%\end{equation}
%where $K$ denotes the number like $k$, but of the rebinned data $\overline{V}_{\rm sum}$, i.e., $t=K \Delta t_{\rm rebin}$.

%%%%%%%%%%%%%%%%%%%%%%%%%%%%%%%%%%%%%%%%%%%%%%%%%%%%%%%%%%%
%%%%%%%%%%%%%%%%%%%%%%%%%%%%%%%%%%%%%%%%%%%%%%%%%%%%%%%%%%%%%%%%%%%%%%%%%%%%%%%%%%%%%%

\subsection{Random Fluctuation}
\label{result-fluctuation}
\begin{figure}[htbp]
 \begin{center}
  \includegraphics[width=86mm]{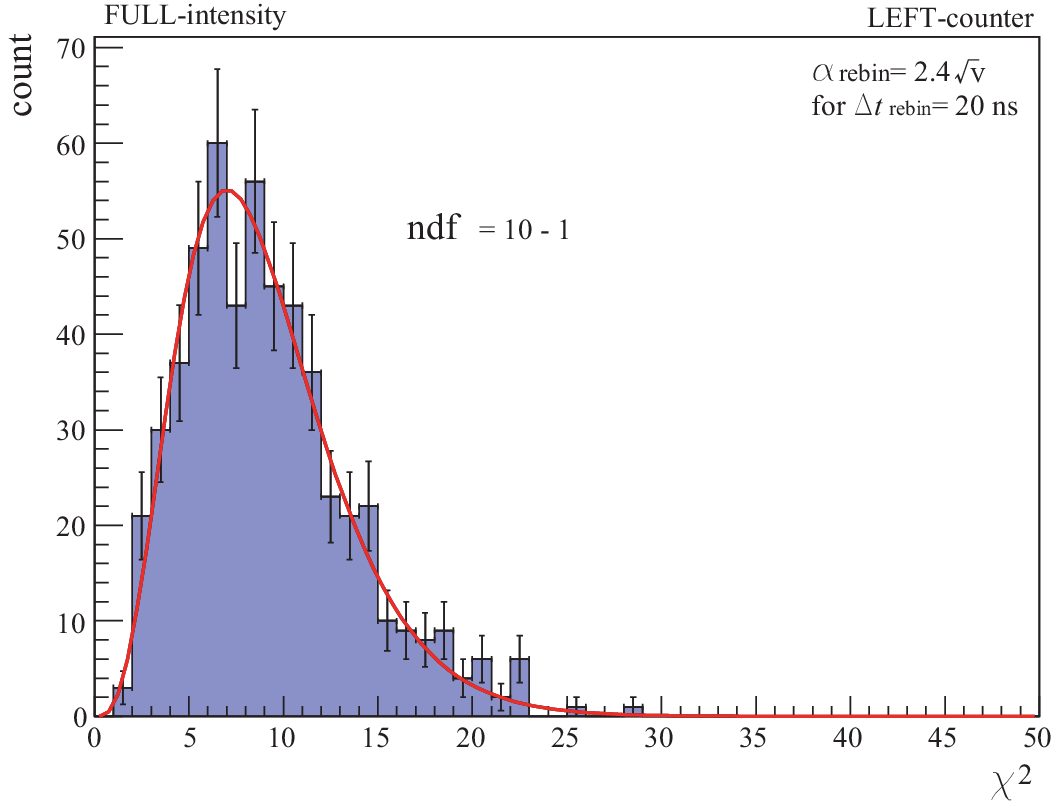}
 \end{center}
 \caption{
Distribution of $\chi^2$ values obtained from the fine-segmented exponential decay curve, as will be described in Section \ref{random} using Fig.\ref{stdfitting}. Ten points were used for the fitting. The error associated with each data point was evaluated using Eq.(\ref{eq-sqrt}), with $\alpha_{\rm rebin}=2.4 \sqrt{\rm V}$. By properly tuning this $\alpha$ value, the obtained distribution demonstrated good agreement with the theoretical $\chi^2$ distribution, with ${\rm ndf}=10-1$, indicated by the solid line.}
 \label{chi2}
\end{figure}

The properties of the random fluctuations in $\overline{V}_{\rm sum}$ were examined using the experimental data, and the corresponding results are shown in Fig.\ref{chi2}. The $\chi^2$ values plotted in this figure were estimated by performing simple exponential fitting across numerous segmented narrow regions (with a width of 200 ns) of the decay curve displayed in Fig.\ref{decay-0G-ch1}. The function adopted for this exponential fitting on $\overline{V}_{\rm sum}$ was
\begin{equation}
V_{\rm model}(t)=A_{\rm V} \; e ^{-t/{\tau_\mu} }.
\end{equation}
%Initially, a two-parameter $(A_{\rm V}, \tau_\mu)$ fitting was performed over a wide region of the decay curve displayed in Fig.\ref{decay-0G-ch1} without segmenting it. 
Initially, a $\chi^2$-fittings with two-parameters $(A_{\rm V}, \tau_\mu)$ was performed over a wide region of the decay curve displayed in Fig.\ref{decay-0G-ch1} without segmenting it. Subsequently, narrow window fittings were performed on segmented decay curves, fixing the value of the lifetime parameter obtained above and setting the scale parameter, $A_{\rm V}$, as the free parameter.  
To perform this $\chi^2$-analysis, we determined the random error $\sigma_{\overline{V}_{\rm sum}}$ by fixing $\alpha$ in Eq.(\ref{eq-sqrt}), which is used in 
\begin{equation}
\chi^2=\sum_{K' \subseteq {\rm fitting\, region}} \left( \frac{\overline{V}_{\rm sum}(K'\Delta t _{\rm rebin})-V_{\rm model}(K'\Delta t _{\rm rebin})}{\sigma_{\overline{V}_{\rm sum}}(K'\Delta t _{\rm rebin})}\right)^2 \;\;.
\end{equation}

The number of independent data points within the segmented 200-ns-fitting region was 10, yielding a degree of freedom of ${\rm ndf}=10-1=9$ for the fitting. The obtained result shown in Fig.\ref{chi2} demonstrates excellent agreement with the expected $\chi^2$ distribution with the parameter $\nu={\rm ndf}$: 
\begin{equation}
P_\nu(\chi^2)=\frac{\left(\chi^2/2\right)^{\nu/2-1} e^{-\chi^2/2}}{2 \Gamma(\nu/2)}.
\end{equation}
This implies that the fluctuation $\sigma_{\overline{V}_{\rm sum}}$ used to estimate $\chi^2$ values obeys a Gaussian distribution, allowing reliable model fitting using conventional $\chi^2$ minimization.

%To obtain the results and perform this $\chi^2$ analysis, we determined the random error $\sigma_{\overline{V}_{\rm sum}}$ in the experimental data $\overline{V}_{\rm sum}$ using
%\begin{equation}
%\chi^2=\sum_{K' \subseteq {\rm fitting\, region}} \left( \frac{\overline{V}_{\rm sum}(K'\Delta t _{\rm rebin})-V_{\rm model}(K'\Delta t _{\rm rebin})}{\sigma_{\overline{V}_{\rm sum}}(K'\Delta t _{\rm rebin})}\right)^2 \;\;.
%\end{equation}
%where $K$ denotes the number like $k$, but of the rebinned data $\overline{V}_{\rm sum}$, i.e., $t=K \Delta t_{\rm rebin}$.

The value of $\alpha$ in Eq.(\ref{eq-sqrt}) was determined as $\alpha_{\rm rebin}=2.4 \sqrt{\rm V}$ to reproduce the $\chi^2$ distribution corresponding to ${\rm ndf}$  by ensuring agreement between the data and theoretical model with ${\rm ndf}=10-1$, as shown in Fig.\ref{chi2}. 
To do that, initially, the $\chi^2$ distribution was generated using a temporary value of $\alpha_{\rm temp.}=1 \sqrt{\rm V}$. Subsequently, the optimized $\alpha$ was approximately obtained using the mean value $\left<\chi^2_{\rm temp.} \right>$ of the $\chi^2$ distribution utilizing $\alpha_{\rm temp.}$, as follows:
\begin{equation}
\alpha=\sqrt{\frac{\left<\chi^2_{\rm temp.}\right>}{\rm ndf}}.
\label{alpha-estimation}
\end{equation} 
This is because $\left<\chi^2_{\rm temp.}\right>/\alpha^2=\left<\chi^2\right>={\rm ndf}$. Thus, we determined $\alpha$ by estimating the fluctuation in the experimental data instead of using Eq.(\ref{alpha-def}). From this analysis, we determined the $\alpha$ factor as $\alpha_{\rm rebin}=2.4\sqrt{\rm V}$ for the rebinned data ($m=10$) by performing fine-tuning in accordance with Fig.\ref{chi2}.

\begin{figure}[htbp]
 \begin{center}
  \includegraphics[width=86mm]{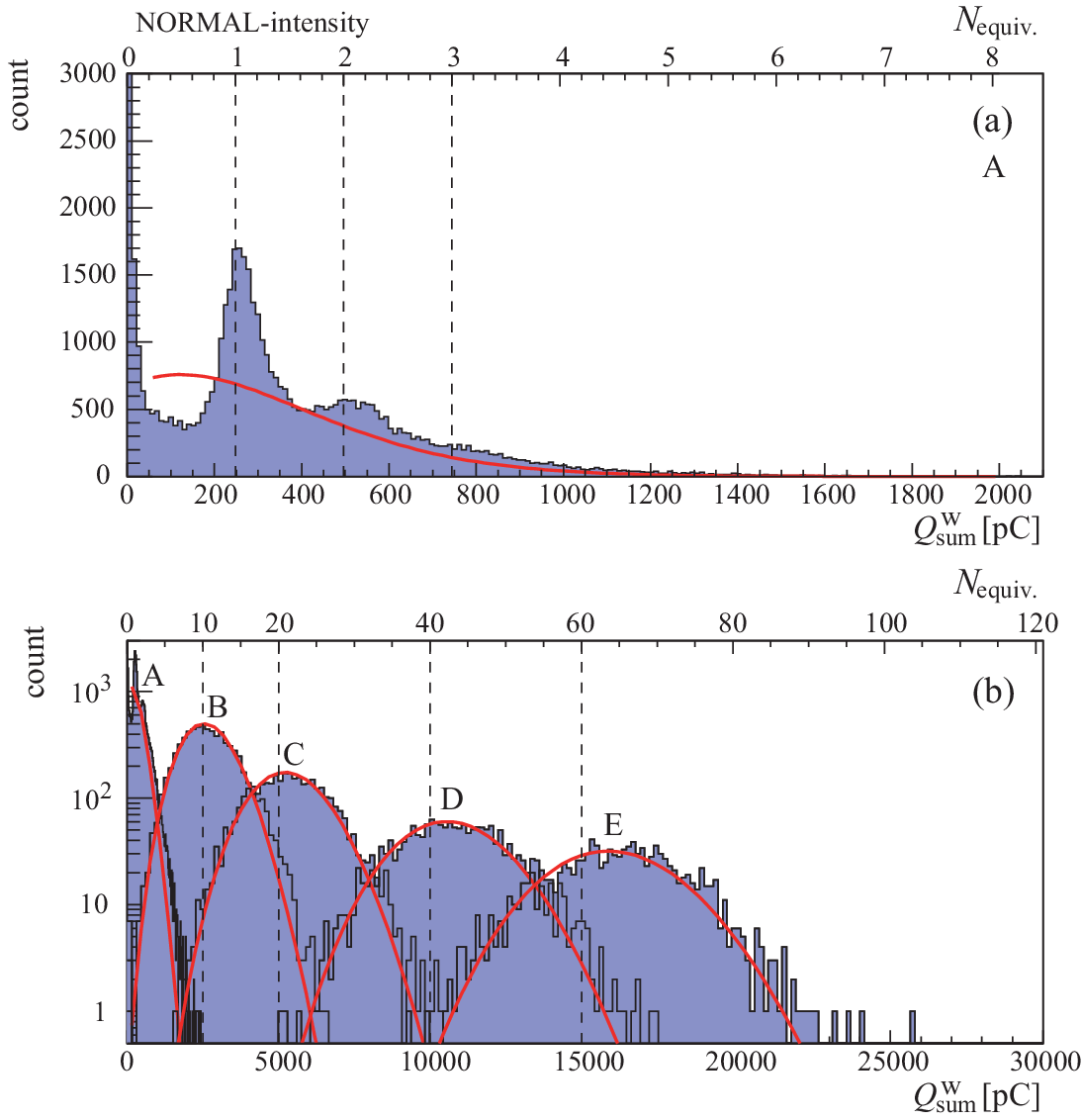}
 \end{center}
 \caption{
Distribution of $Q_{\rm sum}^{\rm w}$  defined in Eq.(\ref{Q_sum-def}). (a) and (b) are similar but show different $Q_{\rm sum}^{\rm w}$ regions. The upper horizontal axis represents the equivalent counting number, $N_{\rm equiv.}$ estimated using the value of $Q_0$. The $Q_{\rm sum}^{\rm w}$ distributions, obtained by tuning the expected value $M$ defined in Eq.(\ref{Poisson-def}) in the $N_{\rm equiv.}$ spectra, demonstrate good agreement with the Poisson distributions. Refer to the main text for details on the $N_{\rm pulse}$ settings for cases A/B/C/D/E defined in Eq.(\ref{Q_sum-def}).
}
 \label{Poisson}
\end{figure}

The abovementioned agreement with the $\chi^2$ distribution suggests the Gaussian properties of the fluctuations in $\overline{V}_{\rm sum}$, which can be directly tested by examining the fluctuating $\overline{V}_{\rm sum}$ distribution itself. We identified a time window $T_{\rm w}$ 
to obtain the time-integrated charge for the $j$-th beam pulse as follows:
\begin{equation}
%Q^{\rm w}_j=\sum_{k}^{k\Delta t \leq T_{\rm w}} q_j(k \Delta t)=-\sum_{k}^{k \Delta t \leq T_{\rm w}} \frac{V_j(k\Delta t)\Delta t}{R_{\rm T}}.    
Q^{\rm w}_j=-\sum_{k=0}^{k \Delta t < T_{\rm w}} \frac{V_j(k\Delta t)\Delta t}{R_{\rm T}}.    
\end{equation}
Following this, the $Q^{\rm w}_j$ distribution over different beam pulses, $j$, was examined, as shown in Fig.\ref{Poisson}. In this figure, cases (A/B/C/D/E) show the results summing $N_{\rm pulse}=$1/10/20/40/60 beam pulses, i.e.,
\begin{equation}
Q_{\rm sum}^{\rm w (A/B/C/D/E)}=\sum_{j=1}^{j \leq 1/10/20/40/50} Q_j^{\rm w}.
\label{Q_sum-def}
\end{equation}
For case (A), the integrated charge distribution displayed peaks at $Q_{\rm sum}^{\rm w (A)} \sim Q_0, 2Q_0,3Q_0$, as presented in Fig.\ref{Poisson}(a). $N_{\rm equiv.}=Q_{\rm w}/Q_0$ (number of event pulses included in $T_{\rm w}$)  was estimated, as plotted along the upper horizontal axis. Indeed, three peaks were observed at the $N_{\rm equiv.}=0,1,2,3$ positions. For cases (B-E), the distribution displayed in Fig.\ref{Poisson}(b) exhibits peaks at $Q_{\rm sum}^{\rm w}\sim 10Q_0,20Q_0, 40Q_0,60Q_0$, corresponding to $N_{\rm equiv.}\sim 10,20,40,60$, respectively.

These results were then compared with the theoretical Poisson distribution. The solid lines in the figures show the corresponding Poisson distributions, with parameter $M$ indicating the expected value of the distribution. Here, we used a continuum function to generate the Poisson distribution by extending the functional form using an integer variable $\kappa$ to that using a real number variable $x$ as follows:
\begin{equation}
P_M(\kappa)=\frac{M^\kappa}{\kappa!} e^{-M} \;\;\;\;(\kappa \in {\rm integer}\geq 0) 
\end{equation}
\begin{equation}
\rightarrow P_M(x)=\frac{M^x}{\Gamma(x+1)} e^{-M}
\;\;\;\;(x \in {\rm real \;number}\geq 0) .
\label{Poisson-def}
\end{equation}
The data and model demonstrated good agreement, indicating that the integrated charge obeys the Poisson distribution. This suggests that the integrated voltage $V_{\rm sum}$ exhibits Poisson fluctuations, which converge to Gaussian fluctuations, corroborating the error estimation accomplished using $\sqrt{\overline{V}_{\rm sum}}$ in Eq.(\ref{eq-sqrt}).

We confirm that the fluctuation properties satisfy the requirement for the application of the conventional $\chi^2$ analysis, which assumes Gaussian fluctuations of the random errors of data points.

%%%%%%%%%%%%%%%%%%%%%%%%%%%%%%%%%%%%%%%%%%%%%%%%%%%%%%%%%%%%%%%%%%%%%%%%%%%%%%%%%%%%%%%%
%%%%%%%%%%%%%%%%%%%%%%%%%%%%%%%%%%%%%%%%%%%%%%%%%%%%%%%%%%%%%%%%%%%%%%%%%%%%%%%%%%%%%%%%
%%%%%%%%%%%%%%%%%%%%%%%%%%%%%%%%%%%%%%%%%%%%%%%%%%%%%%%%%%%%%%%%%%%%%%%%%%%%%%%%%%%%%%%%
\section{Discussion}
\label{Discussion}
%%%%%%%%%%%%%%%%%%%%%%%%%%%%%%%%%%%%%%%%%%%%%%%%%%%%%%%%%%%%%%%%%%%%%%%%%%%%%%%%%%%%%%%%
\subsection{Random Error Estimation}
\label{random}
Using the obtained experimental data, our next analysis focuses on examining the square-root property of Eq.(\ref{eq-sqrt}). As detailed in Section \ref{result-fluctuation}, we performed $\chi^2$ fittings on the fine-segmented regions of the decay spectra to extract the $\chi^2$ distribution displayed in Fig.\ref{chi2}. During this process, the standard deviations, $\sigma_{V_{\rm sum}}$, of the data distributed around the fitted curves were calculated while determining the $\chi^2$ value.

\begin{figure}[htbp]
 \begin{center}
  \includegraphics[width=86mm]{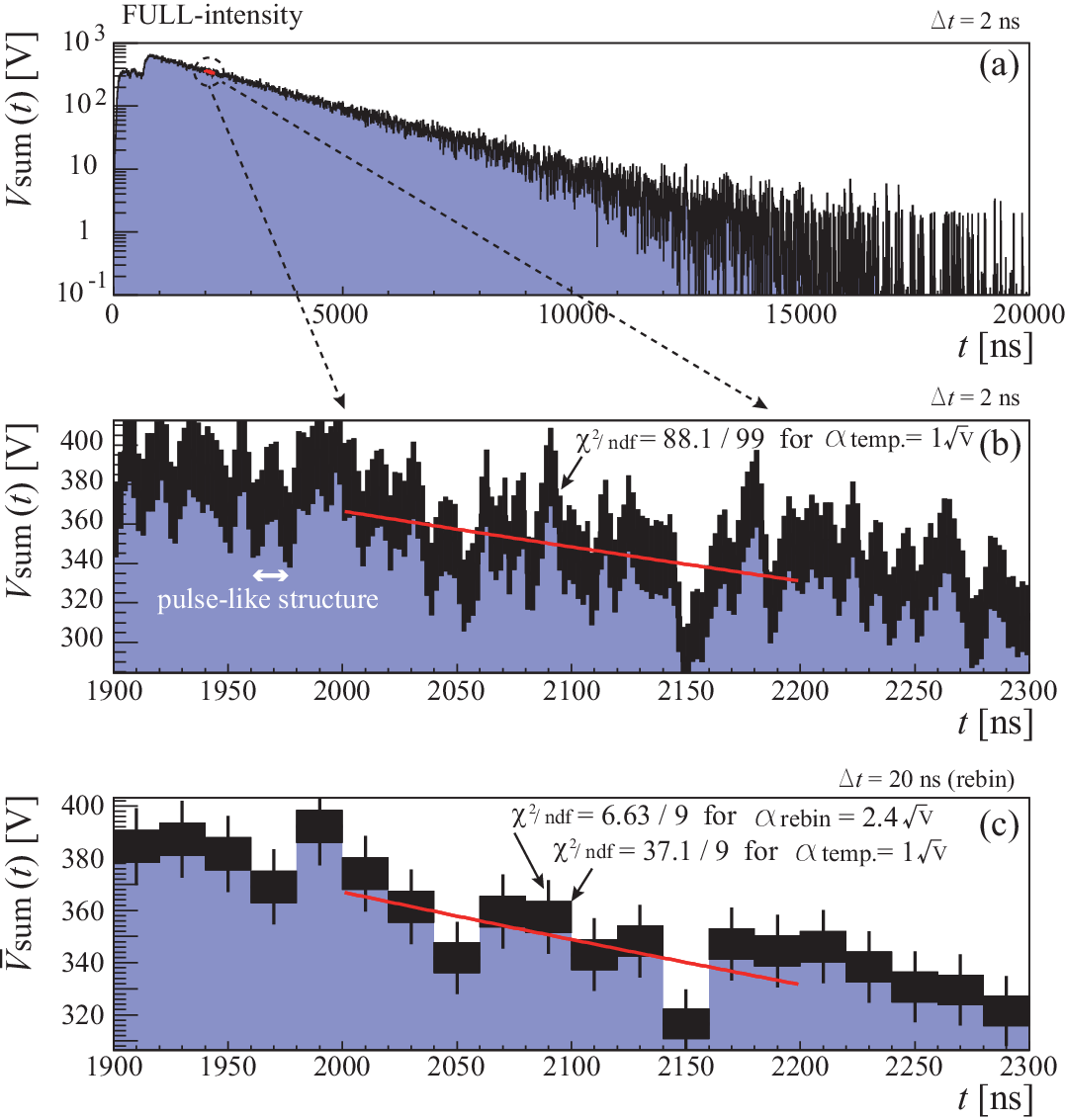}
 \end{center}
 \caption{
Fine-segmented fitting performed using a single exponential function. Solid lines indicate the fitted results. (a) $V_{\rm sum}(t)$ spectrum with $\Delta t=2$ ns, i.e., without rebinning. An example of the fitting region is also shown. (b) Enlarged view of (a) around the typical fitting region as an example. Error bars were estimated using $\alpha_{\rm temp.}=1$. (c) Rebinned results representing $\overline{V}_{\rm sum}(t)$. Average voltages are shown instead of summing $V_{\rm sum}(t)$ across 10 time bins. Two types of error bars are shown, corresponding to $\alpha_{\rm temp.}=1 \sqrt{\rm V}$ (box) and $\alpha_{\rm rebin}=2.4 \sqrt{\rm V}$ (bar), respectively.
}
 \label{stdfitting}
\end{figure}

Fig.\ref{stdfitting} shows the process of segmented fitting. Specifically, we performed simple exponential fitting in narrow segmented regions with a width of 200 ns, as shown in Fig.\ref{stdfitting}(a), and repeated the process by shifting the fitting region to different $t$ values. Fig.\ref{stdfitting}(b,c) presents an enlarged view of the fitting region for (b) the original data with a bin width of $\Delta t = 2 \;{\rm ns}$ and (c) rebinned data with a bin width of $\Delta t_{\rm rebin} = 20 \;{\rm ns}$. Fig.\ref{stdfitting}(a,b) presents the $V_{\rm sum}$ distribution, while Fig.\ref{stdfitting}(c) shows the distribution of $\overline{V}_{\rm sum}$.
%, i.e., averaged $V_{\rm sum}$ value in the rebinning region of $\Delta t_{\rm rebin}=20$ ns.

The numbers of data points included within the regions shown in Fig.\ref{stdfitting}(b) and (c) are $N_{\rm dat}=100$ and $N_{\rm dat}=10$, respectively, owing to the rebinning. The vertical error bars shown in Fig.\ref{stdfitting}(b) were estimated using Eq.(\ref{eq-sqrt}), assuming a tentative value of $\alpha=\alpha_{\rm temp.}=1\times \sqrt{\rm V}$. The $\alpha$ value corresponding to the data in Fig.\ref{stdfitting}(c) ($\alpha_{\rm rebin}$)  was independently evaluated by ensuring that the $\chi^2$ distribution for the rebinned data remained consistent with the theoretically expected $\chi^2$ distribution, as detailed in Section \ref{result-fluctuation}.

Fig.\ref{stdfitting}(c) displays two different error bars. One type was estimated using $\alpha_{\rm temp.}$ represented using black boxes, while the other was estimated using $\alpha_{\rm rebin}$, indicated as solid lines. The fluctuation observed in Fig.\ref{stdfitting}(b) is natural because the fitting result of $\chi^2/{\rm ndf}=88.1/99$ using $\alpha_{\rm temp.}$. This indicates that the tentative selection of $\alpha_{\rm temp.}=1$ was accidentally adequate. However, for the rebinned data displayed in Fig.\ref{stdfitting}(c), we note that the fluctuation around the fitted line becomes excessive if we use the assumed value of $\alpha_{\rm temp.}$. Indeed, $\chi^2/{\rm ndf}=37.05/9$ for $\alpha_{\rm temp.}$ becomes excessively large, while $\chi^2/{\rm ndf}=6.63/9$ for $\alpha_{\rm rebin}$ remains adequate. This implies that the $\alpha$ estimation based on Eq.(\ref{alpha-estimation}) adequately computes the random uncertainty. 
%If statistical errors are treated appropriately, $\alpha$ values must not depend on the binning selection. 
The reason underlying the difference between $\alpha_{\rm temp}$ and $\alpha_{\rm rebin}$ values will be discussed in Section \ref{pulse}.

\begin{figure}[htbp]
 \begin{center}
  \includegraphics[width=80mm]{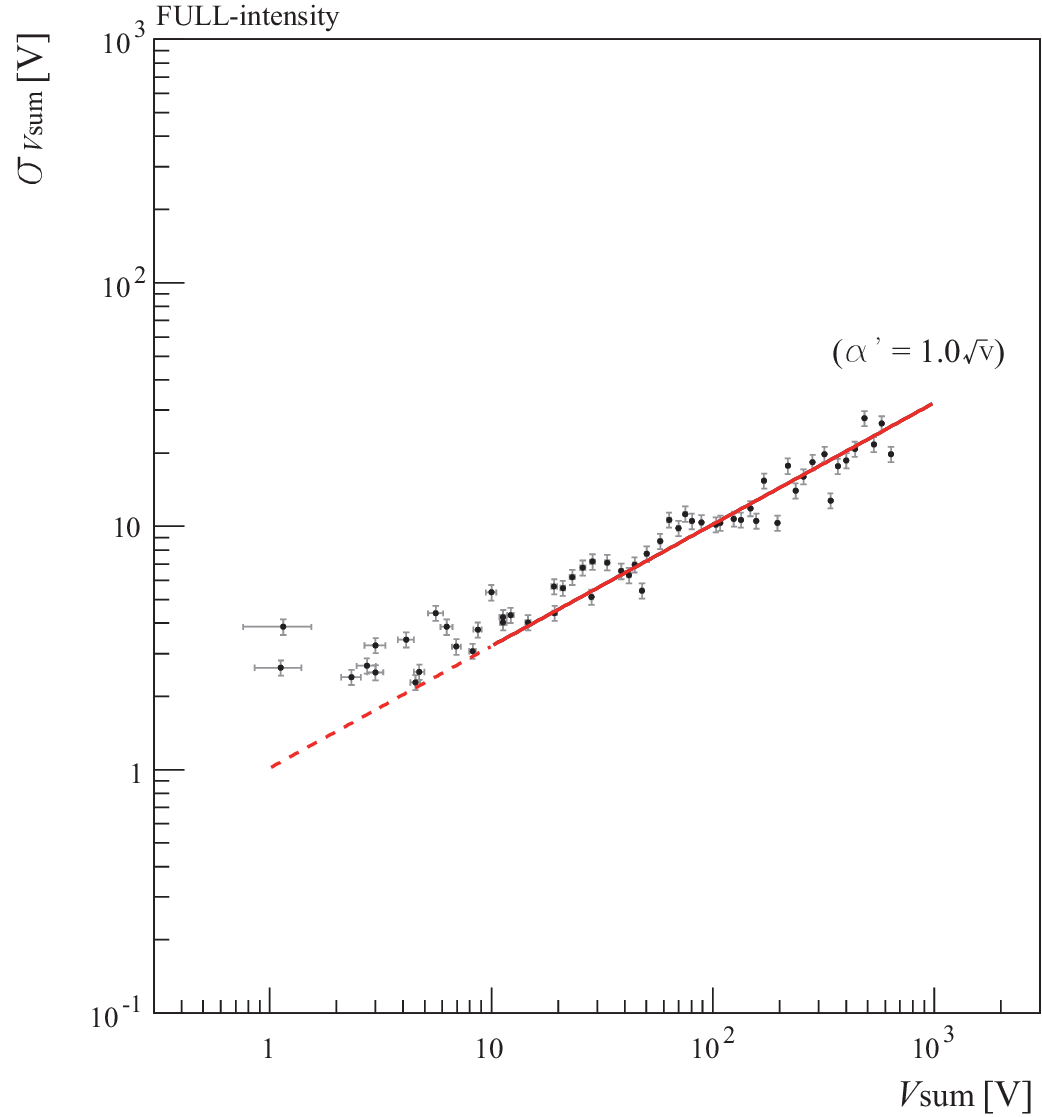}
 \end{center}
 \caption{
$(V_{\rm sum}, \sigma_{V_{\rm sum}})$ correlation fitted using Eq.(\ref{eq-sqrt}). The obtained result is consistent with the square-root property. $\alpha$ value obtained as the fitting result is $\alpha'$. The dotted region was not included in the fitting because individual event pulses dominated the spectrum.
}
 \label{std}
\end{figure}

\begin{figure}[htbp]
 \begin{center}
  \includegraphics[width=80mm]{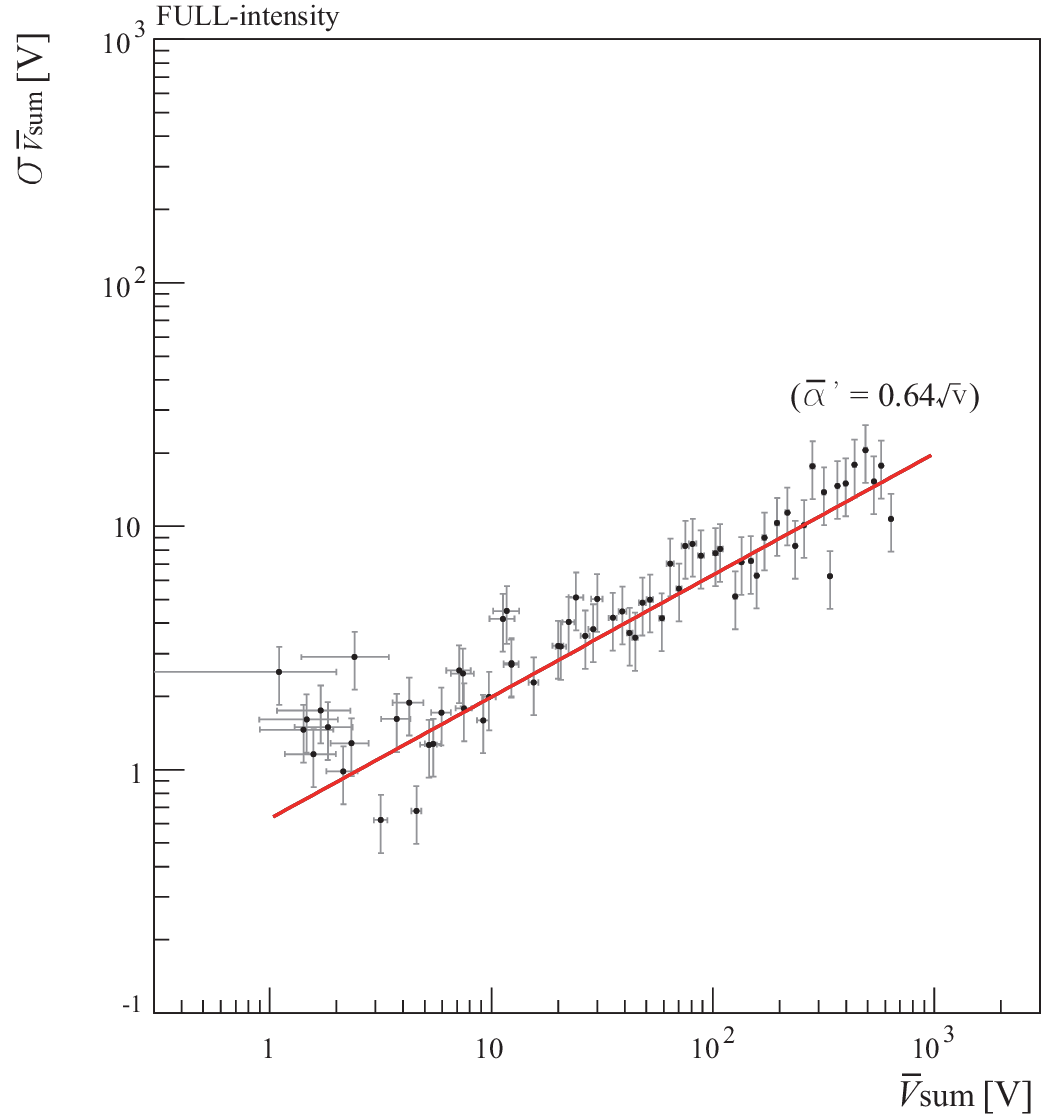}
 \end{center}
 \caption{
$(\overline{V}_{\rm sum}, \sigma_{\overline{V}_{\rm sum}})$ correlation for the rebinned data fitted using Eq.(\ref{eq-sqrt}). The obtained result is consistent with the square-root property. $\alpha$ value obtained as the fitting result is shown as $\overline{\alpha}'$. The entire $\overline{V}_{\rm sum}$ region was involved in the fitting, in contrast to Fig.\ref{std}, owing to the increased statistics resulting from the rebinning of smeared event pulses.
}
\label{std-rebin}
\end{figure}

After performing the fittings, the standard deviations, $\sigma_{V_{\rm sum}}$, were estimated as the fluctuations around the fitted curve, $V_{\rm model}$, for each segmented fitting region, as follows:
\begin{equation}
\sigma_{V_{\rm sum}}^2=\frac{1}{N_{\rm dat}-1}\displaystyle \sum_{k'=0,N_{\rm dat}-1}^{200\;{\rm ns}}  \left( V_{\rm sum}(k'\Delta t)-V_{\rm model}(k'\Delta t) \right) ^2 ,
\label{sigma-fitting}
\end{equation}
where $N_{\rm dat}$ denote the number of data within the 200-ns-wide fitting region. Similarly, the standard deviations for the rebinned data, $\sigma_{\overline{V}_{\rm sum}}$, were also estimated. Subsequently, datasets of $(V_{\rm sum}, \sigma_{V_{\rm sum}})$ and $(\overline{V}_{\rm sum}, \sigma_{\overline{V}_{\rm sum}})$ for the numerous fitting regions were obtained. Figs.\ref{std} and \ref{std-rebin} show their correlations.

By examining these correlations, we can directly test Eq.(\ref{eq-sqrt}). The horizontal errors of $V_{\rm sum}$ shown in Fig.\ref{std} were estimated as their standard errors $\sigma_{V_{\rm sum}}/\sqrt{N_{\rm dat}}$. The vertical errors of $\sigma_{V_{\rm sum}}$ were estimated as the standard errors of standard deviation as $\sigma_{\sigma_{V_{\rm sum}}}=\sigma_{V_{\rm sum}}/\sqrt{2(N_{\rm dat}-1)}$. The errors presented in Fig.\ref{std-rebin} were estimated in the same manner.

The results shown in both figures are consistent with the slope indicating $\sigma \propto V_{\rm sum}^{\;0.5}$. Once again, we successfully confirm the square-root property of Eq.(\ref{eq-sqrt}). We performed fittings on the $(V_{\rm sum},\sigma_{V_{\rm sum}})$ and  $(\overline{V}_{\rm sum},\sigma_{\overline{V}_{\rm sum}})$ correlations by setting $\alpha'$ as the parameter to be optimized, yielding $\alpha'=1.0\,\sqrt{\rm V}$ and $\overline{\alpha'}=0.64\,\sqrt{\rm V}$. The differences in the obtained $\alpha'$ values which are shown in Figs.\ref{std} and \ref{std-rebin} correspond to the differences between $\alpha_{\rm temp}$ and  $\alpha_{\rm rebin}$, as evident from Fig.\ref{stdfitting}. The result $\alpha' \sim \alpha_{\rm temp}$ implies that the tentative assumption of $\alpha_{\rm temp.}=1\,\sqrt{\rm V}$ was accidentally adequate, as observed in Fig.\ref{stdfitting}(b). 

As the random fluctuation is suppressed by considering averaged values, $\sigma_{\overline{V}_{\rm sum}}\sim\sigma_{V_{\rm sum}}/\sqrt{10}$. This occurs because we combine $m=10$ independent data points into one bin during the rebinning process. The obtained fitting results for $\overline{\alpha}'$ are consistent with those for $\alpha_{\rm rebin}$ because $\overline{\alpha}'\sim\alpha_{\rm rebin}/\sqrt{10}$. This difference arises from the distinct definitions of these parameters. For instance, $\alpha_{\rm rebin}$ was used before rebinning, following which $\sigma_{\overline{V}_{\rm sum}}$ was estimated using Eq.(\ref{eq-sqrt}) through error propagation. Meanwhile, $\overline{\alpha}'$ was estimated using the data obtained after rebinning. Either $\overline{\alpha'}$ or $\alpha_{\rm rebin}$ can be used for the analysis; however, estimating $\overline{\alpha}'$ might be simpler.

Thus, this section confirmed the square-root properties for both Figs.\ref{std} and \ref{std-rebin}, directly verifying the relation in Eq.(\ref{eq-sqrt}). The parameter $\alpha$ can be determined either by reproducing the appropriate $\chi^2$ distribution, as shown as $\alpha_{\rm rebin}$ in Fig.\ref{chi2}, or utilizing the $\overline{V}_{\rm sum}$-$\sigma_{\overline{V}_{\rm sum}}$ correlation as $\alpha_{\rm rebin}\sim \overline{\alpha'} \sqrt{10}$, as displayed in Fig.\ref{std-rebin}. In the following study, we will use the $\alpha_{\rm rebin}$ determined from the $\chi^2$ distribution, as described in Section 5.3, to perform $\chi^2$ analysis.
It is because it should be better to directly confirm the $\chi^2$ distribution to perform the $\chi^2$ fitting.

%%%%%%%%%%%%%%%%%%%%%%%%%%%%%%%%%%%%%%%%%%%%%%%%%%%%%%%%%%%%%%%%%%%%%%%%%%%%%%%%%%%%%%%%%
%%%%%%%%%%%%%%%%%%%%%%%%%%%%%%%%%%%%%%%%%%%%%%%%%%%%%%%%%%%%%%%%%%%%%%%%%%%%%%%%%%%%%%

\subsection{Pulse-Like Structure}
\label{pulse}
One may notice that the fluctuation appearing in Fig.\ref{stdfitting}(b) is unnatural. Indeed, the pulse-like structure dose not appear to be random fluctuations. 
%If these fluctuations were, in fact, random and uncorrelated with different time-bin data, the scale parameter $\alpha$ would remain constant regardless of the binning settings, i.e., $\alpha_{\rm temp}$ would match $\alpha_{\rm rebin}$. For conventional pulse-counting TDC histograms, the Poisson error is estimated as $\sigma_N=\alpha_{\rm Poisson} \sqrt{N}=\sqrt{N}$ (where $N$ is the number of counts), implying that $\alpha_{\rm Poisson}=1$ independent of binning settings. 
The presence of pulse-like structures indicates the existence of correlations between adjacent time bins, suggesting that different time-bin data, $V_{\rm sum} (t)$, are not independent.

Similar pulse-like structures were also obtained from simulations, as shown in Fig.\ref{time-sim}, suggesting that bin-to-bin correlated fluctuations do not originate from hardware factors such as baseline fluctuations, bias power instabilities, or timing jitter of the trigger signal but are rather inherent to the mathematical method.

As we saw in Fig.\ref{stdfitting}(c), the fitting $\chi^2$ value becomes excessively large when using $\alpha_{\rm temp.}$ to estimate the errors of the rebinned data. Adopting an alternative approach, we now assume that the estimated error bars displayed in Fig.\ref{stdfitting}(b) must be excessively small, i.e., $\alpha_{\rm temp.}$ must be too small to obtain an adequate value. Indeed, a larger value of $\alpha_{\rm rebin}$ well reproduces the actual fluctuation. Thus, owing to bin-to-bin correlations, the fluctuation was estimated to be small. Given that the observed pulse-like structures demonstrate similar values across adjacent bins, the standard deviation $\sigma_{V_{\rm sum}}$ can be interpreted to decrease.

This unnatural correlation can be understood based on the relation between the event-pulse width $T_{\rm pulse} \sim 10$ ns and the sampling time bin $\Delta t=2$ ns as $T_{\rm pulse} >\Delta t$. Owing to this wide pulse width, charges originating from the same radiation event are distributed across multiple ($\sim T_{\rm pulse}/\Delta t$) sampling time bins, leading to the persistence of the bin-to-bin correlation within a time scale of $T_{\rm pulse}$.
As outlined in Section \ref{result-fluctuation}, although the voltage fluctuation may obey Poisson statistics, the data points cannot be treated as independent samples. Independence between data points is essential for performing $\chi^2$ analysis. Consequently, the $\chi^2$ values obtained without accounting for this behavior are unreliable.

Additionally, the fine time structure apparent within the time scale of $T_{\rm pulse}$ does not accurately represent the time structure of the real event rate. To avoid such confusion, we must ignore the fine time structure within $T_{\rm pulse}$. Consequently, we applied a 10-bin rebinning procedure, wherein the time width was selected to be $\Delta t_{\rm rebin} \geq T_{\rm pulse}$, effectively filtering out high-frequency components. 
Figs.\ref{decay-0G-ch1} and \ref{decay-123G-ch1} show the rebinned ($\Delta t_{\rm rebin} =20\; {\rm ns}$) $\overline{V}_{\rm sum}$ spectra facilitating reliable $\chi^2$ fitting analysis. Conversely, Fig.\ref{decay-0G-ch1-tail} plots the corresponding data without re-binning ($\Delta t =2\; {\rm ns}$) to show the single pulse structure in the tail region.

Thus, $V_{\rm sum}$ is proportional to the positron flux, and its random fluctuations obey a Poisson distribution. However, for reliable $\chi^2$ analysis, time bins must be rebinned over $T_{\rm pulse}$.

%%%%%%%%%%%%%%%%%%%%%%%%%%%%%%%%%%%%%%%%%%%%%%%%%%%%%%%%%%%%%%%%%%%%%%%%%
%%%%%%%%%%%%%%%%%%%%%%%%%%%%%%%%%%%%%%%%%%%%%%%%%%%%%%%%%%%%%%%%%%%%%%%%%%%%%%%%%%%%%%

\subsection{Saturation}
\label{saturation}
The primary objective of this study is to overcome counting saturation encountered during high-event-rate measurements. Notably, the results shown in Fig.\ref{decay-0G-ch1} can be fitted using a single exponential decay curve, indicating negligible saturation effects. The maximum counting rate of the LEFT counter was $n_{\rm LEFT}\sim5$ Mcps, thus establishing a reliable upper limit for this method. 
Fig.\ref{decay-0G-ch1} indicates a maximum hitting rate of 20 Mhps (Mega hit/s), deduced from the beam intensity and detector acceptance, indicating differences in detector efficiency.

\begin{figure}[htbp]
 \begin{center}
  \includegraphics[width=86mm]{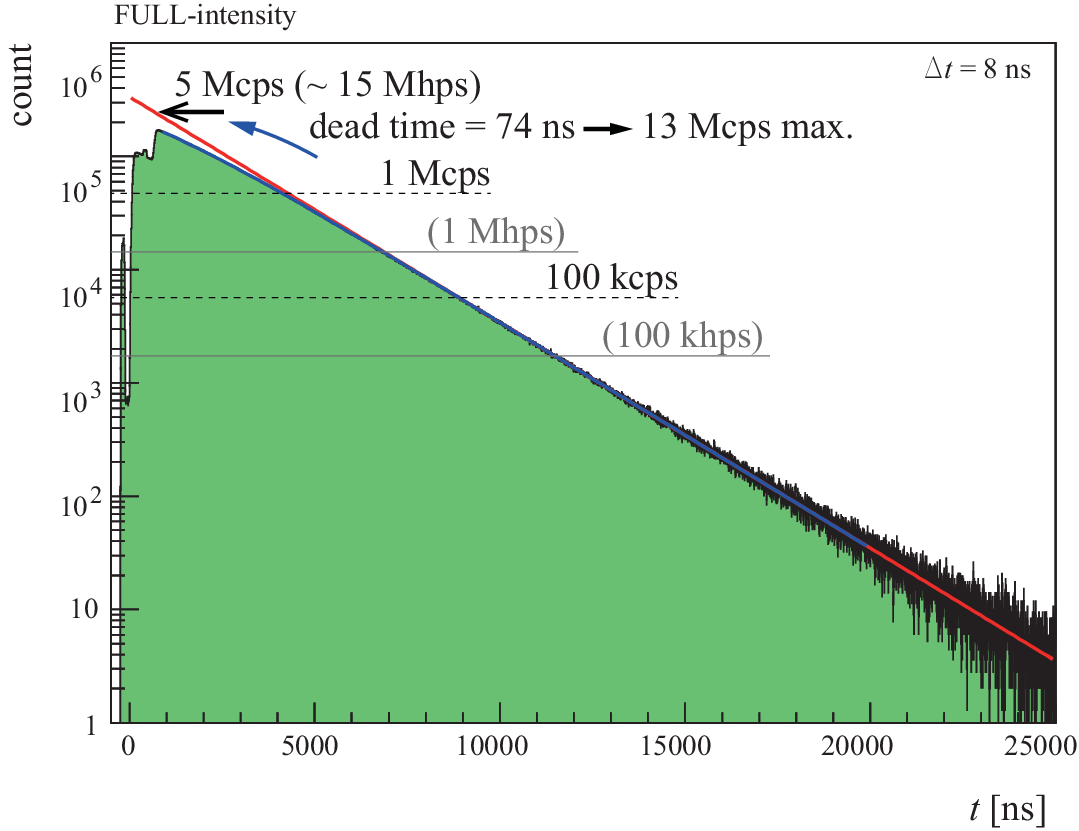}
 \end{center}
\caption{
Decay curve spectrum recorded by Kalliope under FULL-intensity beam exposure. ``hps' indicates hits per second, evaluated based on the single hit rate of the scintillators. ``cps'' indicates accepted coincidence counts between the inner and outer layers per second.
Their difference comes from coincidence efficiency.
}
 \label{Kalliope}
\end{figure}

To benchmark this rate capability, we compared it with those of other devices using voltage discrimination. Fig.\ref{Kalliope} shows the TDC spectra $N(t)$ recorded by Kalliope, with a solid angle per channel of $\Omega_{\rm Kalliope}/\Omega_{\rm LEFT}=0.86$ for the LEFT counter. These spectra were recorded for the same muon beam over $N_{\rm pulse}$ as the data shown in Fig.\ref{decay-0G-ch1} under FULL-intensity beam exposure. Although each segment channel of Kalliope has $\Omega_{\rm Kalliope}$ acceptance, the statistics shown in Fig.\ref{Kalliope} include contributions from all 640 coincidence-channels, justifying the higher statistics in Fig.\ref{Kalliope} compared to those in Fig.\ref{decay-0G-ch1}.

Notably, Kalliope exhibits apparent event-number saturation at a counting rate of $\sim$1 Mcps per channel. The expected hitting rates shown in the figure were obtained by estimating the detection efficiency based on the coincidence rate between two layers (refer to Appendix \ref{beam} for details), thus concluding that our new system is more robust in terms of rate capabilities.

The FORWARD counter was instrumental in determining the maximum event rate of our system, primarily influenced by the substantial background radiation from the beam upstream. The primary types of background radiation expected at the FORWARD counter included positrons emitted from muons decaying upstream of the stopper. Consequently, their decay curves mirrored those derived from the muons stopped at the stopper. The LEFT and RIGHT counters were almost insensitive to these radiation types owing to their setting angles and the use of radiation protection shields upstream of the detector.

\begin{figure}[htbp]
 \begin{center}
  \includegraphics[width=86mm]{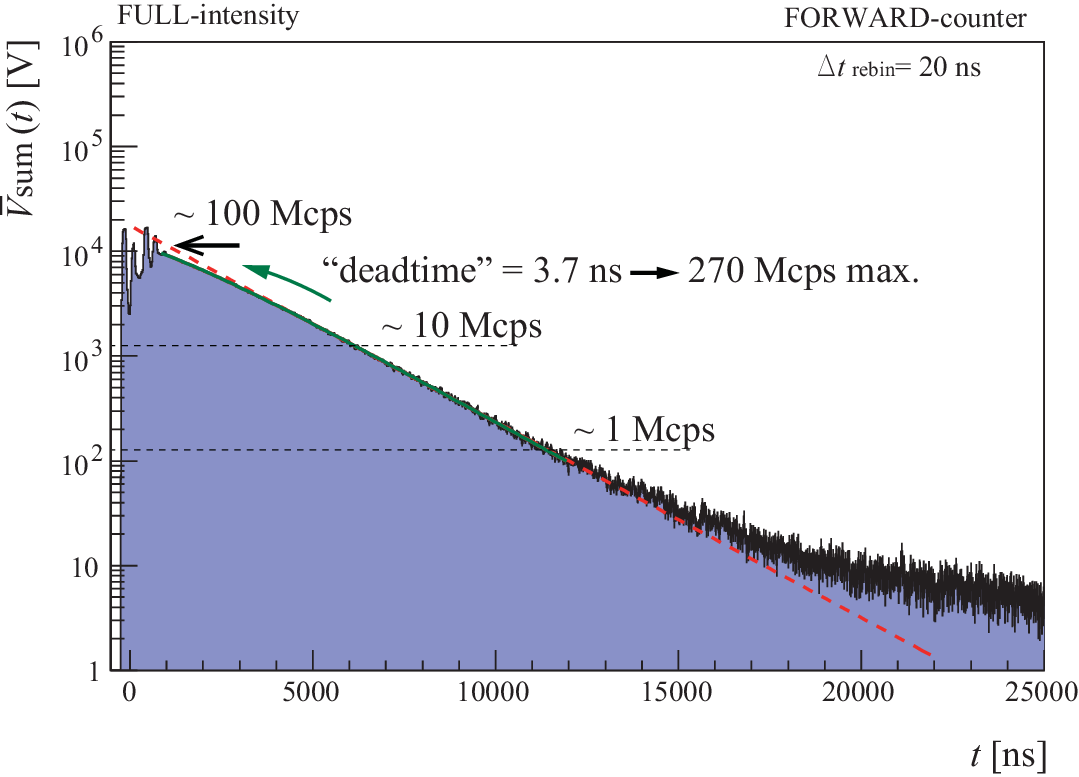}
 \end{center}
 \caption{
Same as Fig.\ref{decay-0G-ch1}, but related to the FORWARD counter. Deviations from the single exponential fitted line are apparent. Based on these deviations, the effective deadtime parameter, $d$, was estimated using Eq.(\ref{n-deadtime}).
There is a component with a time constant of $\sim 10 \,\upmu{\rm s}$. Although the origin of this long tail component remains unidentified, muon decay can be effectively ruled out as a potential source.
}
 \label{decay-0G-ch7}
\end{figure}

\begin{figure}[htbp]
 \begin{center}
  \includegraphics[width=86mm]{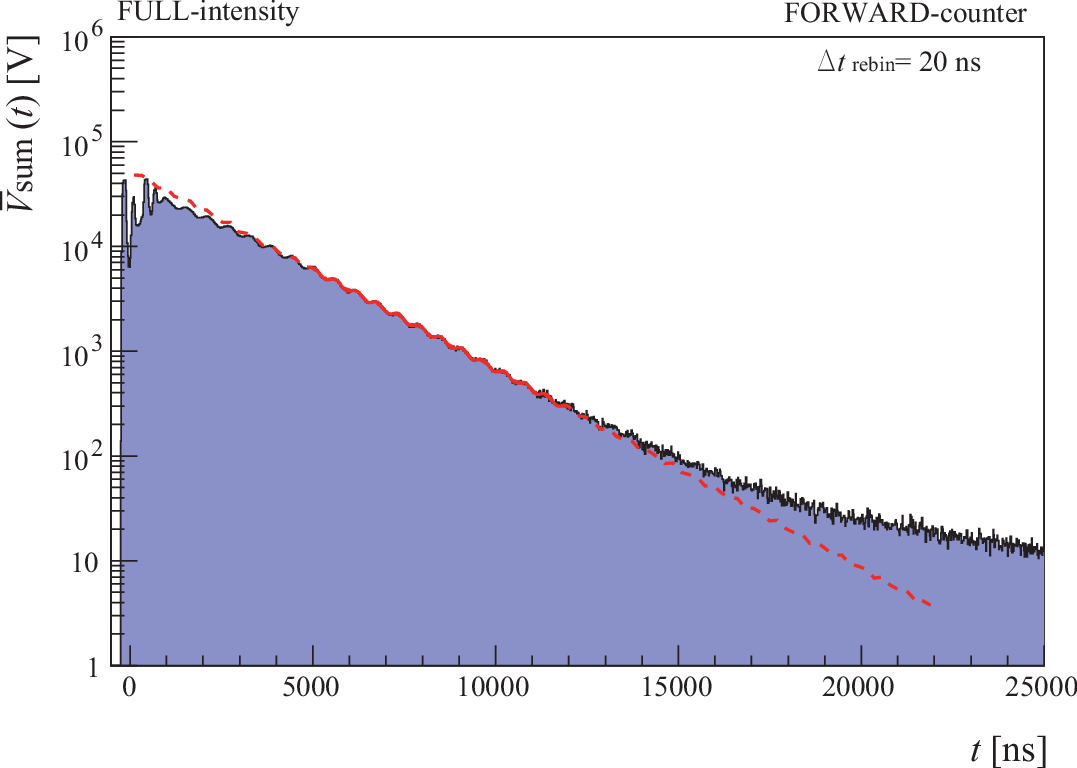}
 \end{center}
 \caption{
Same as Fig.\ref{decay-0G-ch7}, but a transverse magnetic field was applied to generate spin precession for the displayed data. The solid line indicates the fitting result obtained using the exponential function, including spin precession, illustrated as the oscillation pattern around the exponential decay curve. 
}
 \label{decay-123G-ch7}
\end{figure}

The results shown in Figs.\ref{decay-0G-ch7} and \ref{decay-123G-ch7} indicate a small saturation at $\sim$100 Mcps. Fig.\ref{pulse-3} corresponds to this situation. 
The counting rates for the FORWARD counter shown in Fig.\ref{decay-0G-ch7} were evaluated by using the same $Q_0$ value determined for the LEFT counter with Eq.(\ref{eta}).
%This ``$Q_0$-interpretation'' was necessary considering the wide variety of energy loss due to various types of radiation at the FORWARD counter.  
This ``$Q_0$-interpretation'' was necessary considering the wide variety of energy loss due to various types of radiation at the FORWARD counter to interpret the PMT current as equivalent positron-event numbers.
In this sense, this result only demonstrates the current driving capabilities of the PMT, not the actual radiation rate at the FORWARD counter.

The spin precession amplitude displayed in Fig.\ref{decay-123G-ch7} is smaller than that in Fig.\ref{decay-123G-ch1}. This is because most positrons hitting the FORWARD counter do not originate from the stopper, where the transverse magnetic field was applied.

Ideally, the current-readout method must not involve any real ``deadtime." Potential reasons for the saturation may include the saturation of PMTs or the voltage cut-off at 2 {\rm V}, as shown in Fig.\ref{pulse-3}. 
% mod ver3
%Although this saturation effect does not originate from the deadtime of the electronics equipment, the data can be well reproduced using the constant deadtime function as follows:
%\begin{equation}
%n_{\rm accept}=\frac{1}{1/n_{\rm request}+d},
%\label{deadtime}
%\end{equation}
%where $n_{\rm accept \,(request)}$ denotes the accept (request) rate, and $d$ represents the deadtime. 
Although this saturation effect does not originate from the deadtime of the electronics equipment, the data can be well reproduced using the constant deadtime function Eq.(\ref{n-deadtime}).
$d=74 \;{\rm ns}$ is obtained for the Kalliope, as shown in Fig.\ref{Kalliope}. Meanwhile, $d=3.7 \;{\rm ns}$ is obtained for the  FORWARD counter, as shown in Fig.\ref{decay-0G-ch7}.
It was obtained in $\overline{V}_{\rm sum} (t)$ fitting with Eqs.(\ref{n-deadtime}) and (\ref{eta}).

\begin{figure}[htbp]
 \begin{center}
  \includegraphics[width=86mm]{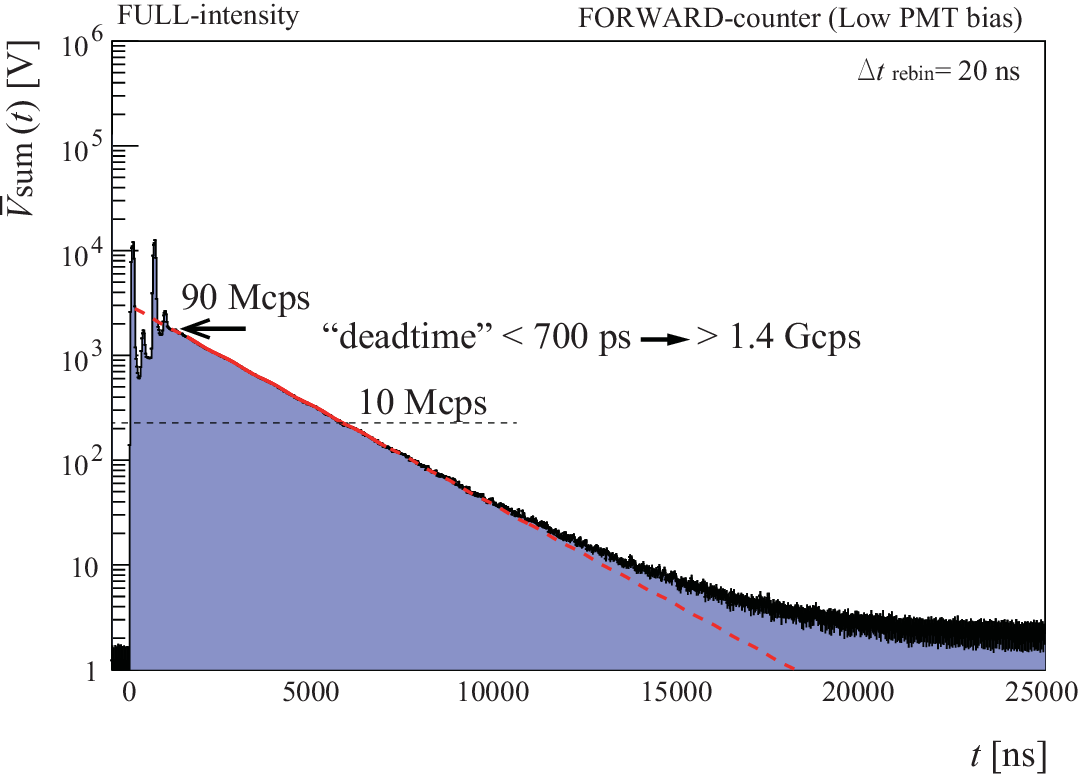}
 \end{center}
 \caption{
Same as Fig.\ref{decay-0G-ch7} but corresponding to the $\mu$LV Run-1 experiment, which involved a lower PMT bias voltage and an voltage attenuator. No saturation effect is apparent. 
%The offset component should be due to the inadequate electric setting of the PMT.
}
 \label{Run1}
\end{figure}

To examine the origin of the observed saturation, we performed another measurement through a different beam experiment, $\mu$LV-Run1, considering the inadequate gain setting of the PMTs for the data shown in Fig.\ref{decay-0G-ch7}. The $\mu$LV-Run1 experiment was performed after the first $\mu$LV-Run0 experiment, under nearly identical conditions. In particular, for the new experiment, we performed a similar test measurement by setting a lower bias voltage from 1800 V to 1400 V and using a voltage attenuator to reduce the pulse height to $\sim$ 200 mV. Again, the ``$Q_0$-interpretation'' was applied using the Run0 value. The corresponding result is shown in Fig.\ref{Run1}.

Although the maximum rate was not as high as that displayed in Fig.\ref{decay-0G-ch7}, no visible saturation could be observed up to at least 90 Mcps. The slight saturation apparent in Fig.\ref{decay-0G-ch7} can be mitigated by reducing the bias voltage of the PMT and/or attenuating the output signals.
$d <$ 700 ps was obtained by a fitting using Eq.(\ref{n-deadtime}) for this case.

\begin{figure}[htbp]
 \begin{center}
  \includegraphics[width=86mm]{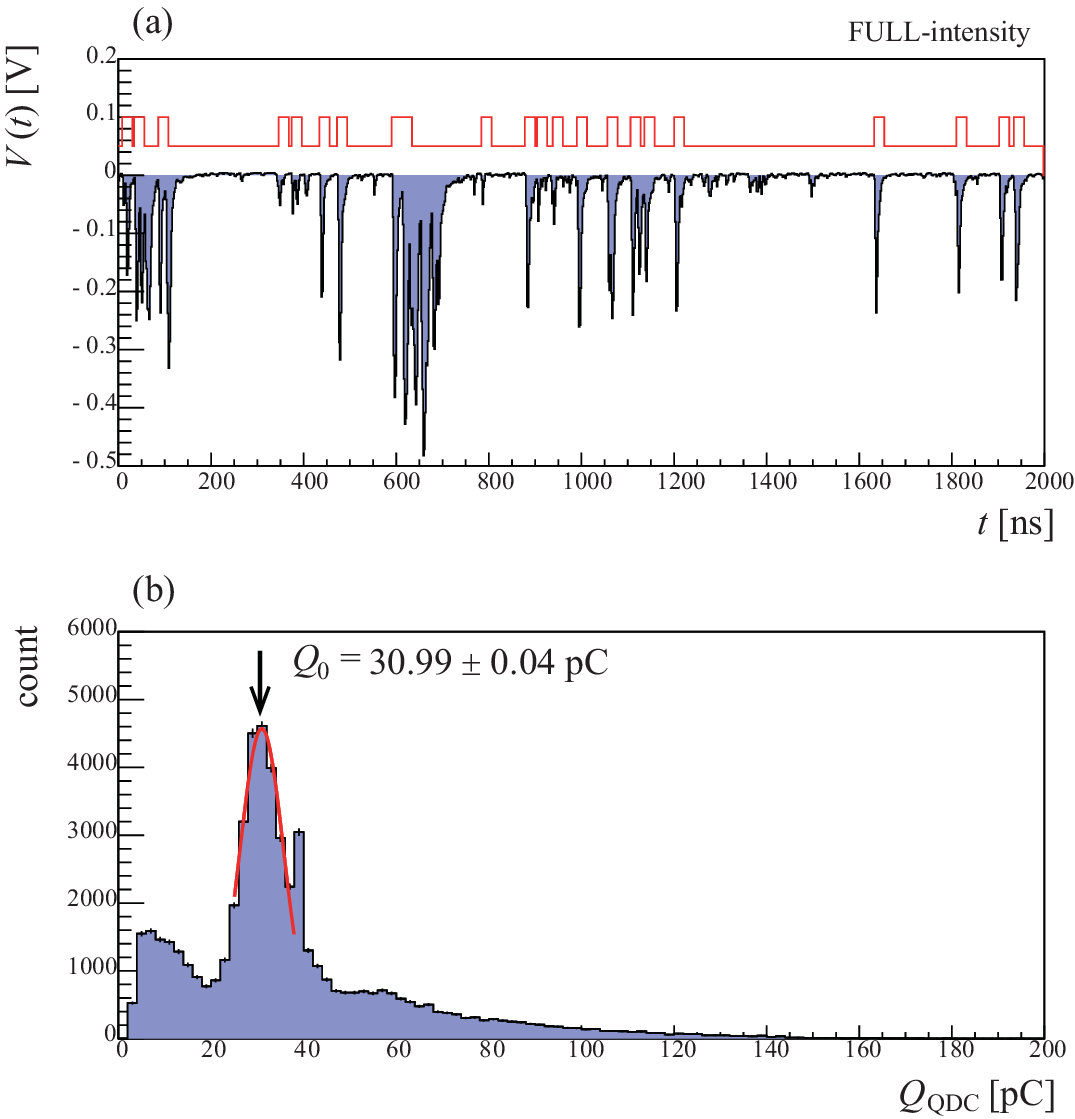}
 \end{center}
 \caption{
Same as Fig.\ref{pulse-3} but corresponding to the $\mu$LV Run-1 experiment, involving a lower PMT bias voltage and a voltage attenuator. (a) The voltage saturation at 2 V is avoided. (b) Smaller $Q_0$ value is confirmed.
}
 \label{pulse-Run1}
\end{figure}

Using Eq.(\ref{n-deadtime}), we can estimate the maximum event rate that can be counted. If we define the maximum acceptable counting rate as the rate when 50\% of the hit events are rejected, i.e., $n_{\rm accept}=n_{\rm request}/2$, the maximum rate $n_{\rm request}^{\rm max}$ can be estimated as follows:
\begin{equation}
n_{\rm accept}=n_{\rm request}^{\rm max}/2=\frac{1}{1/n_{\rm request}^{\rm max}+d} \;\;, \nonumber 
\end{equation}
which yields
\begin{equation}
n_{\rm request}^{\rm max}=1/d \;\;.
\label{deadtime-max}
\end{equation}
Then, $n_{\rm request}^{\rm max}=1/(3.7\; {\rm ns}) \rightarrow 270 \;{\rm Mcps}$ is expected in Fig.\ref{decay-0G-ch7} obtained in the setting of $\mu$LV-Run0.
From the result in Fig.\ref{Run1} for  $\mu$LV-Run1 by reducing the PMT's HV setting, $n_{\rm request}^{\rm max} > 1/(700 \; {\rm ps}) \rightarrow 1.4 \;{\rm Gcps}$ was obtained.
Fig.\ref{pulse-Run1} shows the signal pulses and the QDC spectrum, showing no voltage saturation and a small $Q_0$ value due to a lower HV setting.

\begin{figure}[htbp]
 \begin{center}
  \includegraphics[width=86mm]{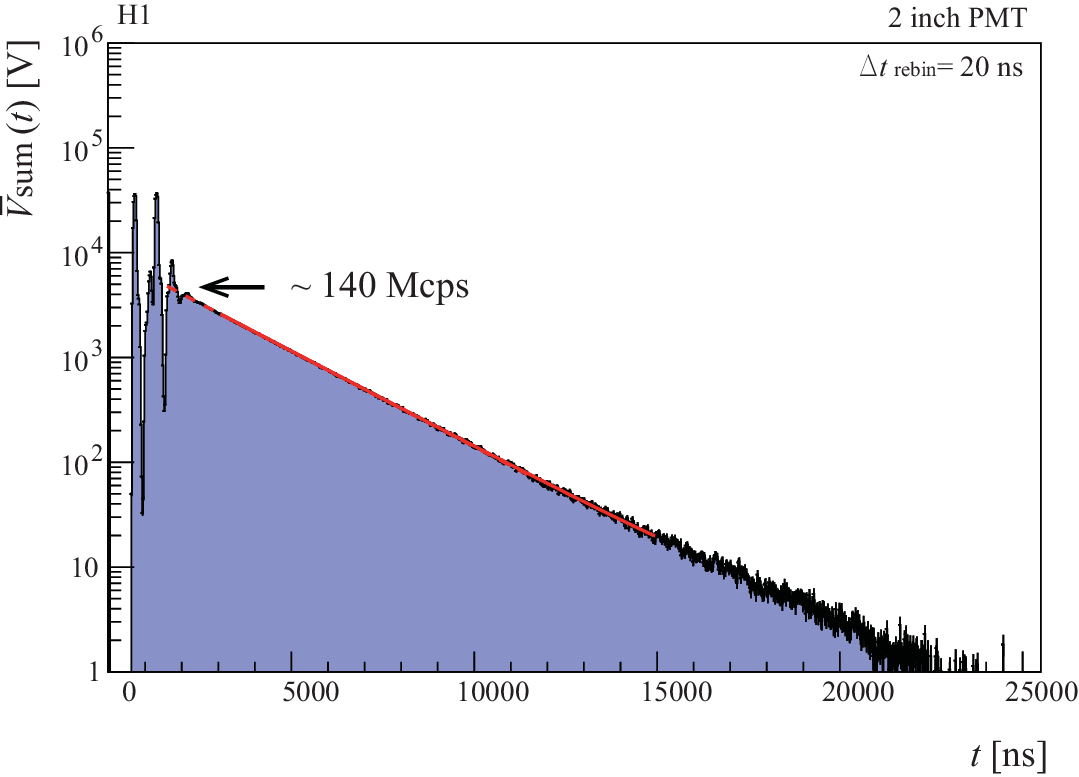
}
 \end{center}
 \caption{
Same as Fig.\ref{decay-0G-ch7} but corresponding to the measurement at H1 area. No saturation effect is apparent. 
}
 \label{H1}
\end{figure}

Finally, we performed another additional measurement at the H1 area \cite{Yamazaki_2023}, where a much stronger beam was available than the D1 area at J-PARC MLF.
We set the LEFT counter using a PMT (Hamamatsu H7195, plus 51mm diameter,  thickness of 0.5 mm BC408 plastic scintillator), 600 mm apart from the center of the aluminum beam stopper.
This time, radiation hitting the counter was limited to positrons emitted from the stopper; then, the counting rate was directly estimated using its own $Q_0$ value without relying on the ``$Q_0$-interpretation'' as in Figs.\ref{decay-0G-ch7} and \ref{H1} removing the ambiguities on this interpretation.
Its result is shown in Fig.\ref{H1}, indicating 140 Mcps counting rate at the maximum with negligible saturation, promising $n_{\rm request}^{\rm max}>1 \;{\rm Gcps}$.

In summary, the proposed method is anticipated to allow $>$1 Gcps measurements if the output voltage is within the digitizer's acceptable range and if the PMT does not saturate.

%%%%%%%%%%%%%%%%%%%%%%%%%%%%%%%%%%%%%%%%%%%%%%%%%%%%%%%%%%%%%%%%%%%%%%%%%%%%%
%%%%%%%%%%%%%%%%%%%%%%%%%%%%%%%%%%%%%%%%%%%%%%%%%%%%%%%%%%%%%%%%%%%%%%%%%%%%%%%%%%%%%%

\subsection{PMT's Saturation}
\begin{figure}[htbp]
 \begin{center}
  \includegraphics[width=86mm]{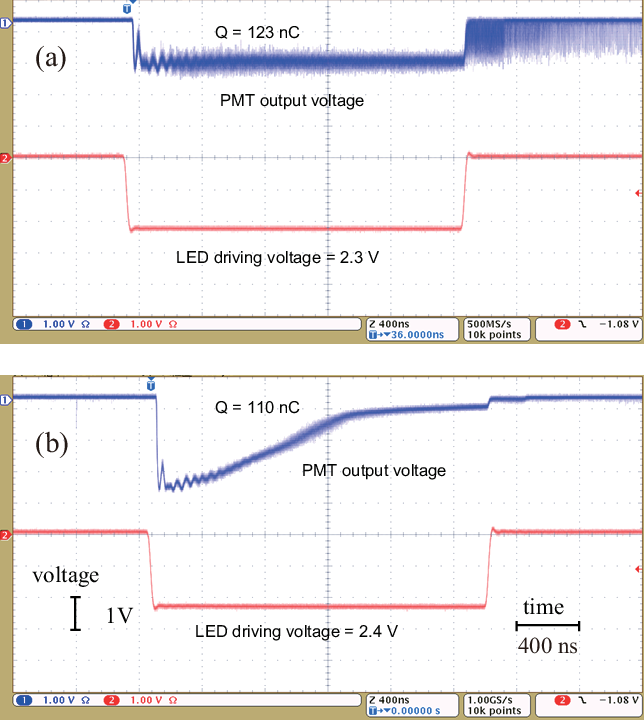} 
 \end{center}
 \caption{
Typical profile recorded by an oscilloscope during the PMT-illuminating test conducted using an LED. Rectangular voltage pulses drove the LED. Total charges $Q$ are presented. These were estimated from the pulse current flowing from the PMT.
}
 \label{PMT-saturation}
\end{figure}

To test the saturation of the PMTs, we performed a dedicated offline measurement using a light-emitting diode (LED) to illuminate the photocathode. Specifically, a blue LED was placed within the LEFT counter to emit visible light. The input voltage for the LED driver was a rectangular pulse with a width of 2 $\upmu$s and height of $V_{\rm LED}=1.2$ V, as shown in Fig.\ref{PMT-saturation}. By adjusting $V_{\rm LED}$, we could vary the amount of light hitting the PMT. The output signal of the PMT subjected to a voltage bias of 1800 V observed using $R_{\rm T}=50\, \Omega$ termination is also shown, and its shape differs from that of the input light signal.
The anode of the PMT was terminated using an internal resistor with a resistance of $50\; \Omega$. Notably, the charge measured in the previous experiments corresponds to the current flowing through $R_{\rm T}$. To estimate the total charge generated within the PMT, we must consider the total resistance as 25 $\Omega$.

From the obtained result, the maximum charge of the PMT over a short period on the order of $\sim \,\upmu$s is $Q_{\rm saturation}\sim$ 100 nC. For larger amounts of charge, the signal output must be suppressed. The saturated event number is roughly estimated as $n_{\rm saturation}=Q_{\rm saturation}/(Q_0\times 50\Omega/25\Omega) \sim 200$ events/beam pulse if photoelectrons arrive at the PMT as a continuous current.
In practical situations, input photoelectrons arrive in the form of pulse-like currents. Therefore, the saturation effect may be mitigated over the event-off timing. The small saturation of $V_{\rm sum} $ observed in Fig.\ref{decay-0G-ch7} might be partially attributed to this saturation of the PMT.

When using a large number of PMTs, the power stability of high-voltage power supplies must be considered. One approach to avoid the saturation and bias power instabilities of PMTs involves reducing the bias voltage. However, reducing the bias voltage or usage of voltage attenuation may exacerbate output charge fluctuations, i.e., $\sigma_{\beta}$. 
%Consequently, voltage attenuation must be performed before processing the digitizer input.

%%%%%%%%%%%%%%%%%%%%%%%%%%%%%%%%%%%%%%%%%%%%%%%%%%%%%%%%%%%%%%%%%%%%%%%%%%%%%%%%%%%%%%
%%%%%%%%%%%%%%%%%%%%%%%%%%%%%%%%%%%%%%%%%%%%%%%%%%%%%%%%%%%%%%%%%%%%%%%%%%%%%%%%%%%
\subsection{Coincidence}
Our findings indicate that the current-readout method can be effectively applied at a high event rates featuring severe pileups. This approach seems impossible to make coincidence measurements between different detector channels, as it does not involve individual pulse identification. That is why we can avoid the pile-up effects.

However, the current-readout method can still perform coincidence (AND) operations. In the conventional pulse-counting method, the AND operation is based on two-adic (binary) logic circuits.
\begin{eqnarray}
{\rm TRUE} \cap {\rm TRUE} &=& 1 \times 1 = 1 \nonumber \\
{\rm TRUE} \cap {\rm FALSE} &=& 1 \times 0 = 0 \nonumber \\
{\rm FALSE} \cap {\rm FALSE} &=& 0 \times 0 = 0.
\label{BOOL}
\end{eqnarray}
In this case, voltage discrimination is adopted to identify whether the state has a value of one or zero.

The current-readout method, however, does not use voltage discrimination; consequently, we cannot identify the value of the state into one or zero. However, distinguishing between TRUE or FALSE is not always necessary. Notably, the output voltage $V(t)$ represents the event flux observed at the detectors, implying that $V(t)$ can be interpreted to be proportional to the ``state.'' This means that instead of viewing the problem from the perspective of two-adic Boolean algebra, we can regard $V(t)$ as the logic signal, described either as in analog form, as a real number, or as an $n$-adic number. Digitization may be applied at the final stage; however, it does not have to be applied in the first stage.

Based on this, we can extend Eq.(\ref{BOOL}) from two-adic Boolean algebra to cover operations between two real numbers as follows:
\begin{eqnarray}
V_{\rm AND}(t)&=&V_1(t) \times V_2(t) \nonumber \\
V_{\rm OR}(t)&=&V_1(t) + V_2(t).
\label{real}
\end{eqnarray}
Eq.(\ref{real}) includes Eq.(\ref{BOOL}) as a special case of $V= 0\; {\rm or}\; 1$. This extension implies a shift in our perspective from viewing TRUE or FALSE as a binary black-and-white image to considering it as a grayscale image. In this renewed framework, voltage represents TRUENESS, defined as the reliability of being regarded as TRUE.

\begin{figure}[htbp]
 \begin{center}
  \includegraphics[width=86mm]{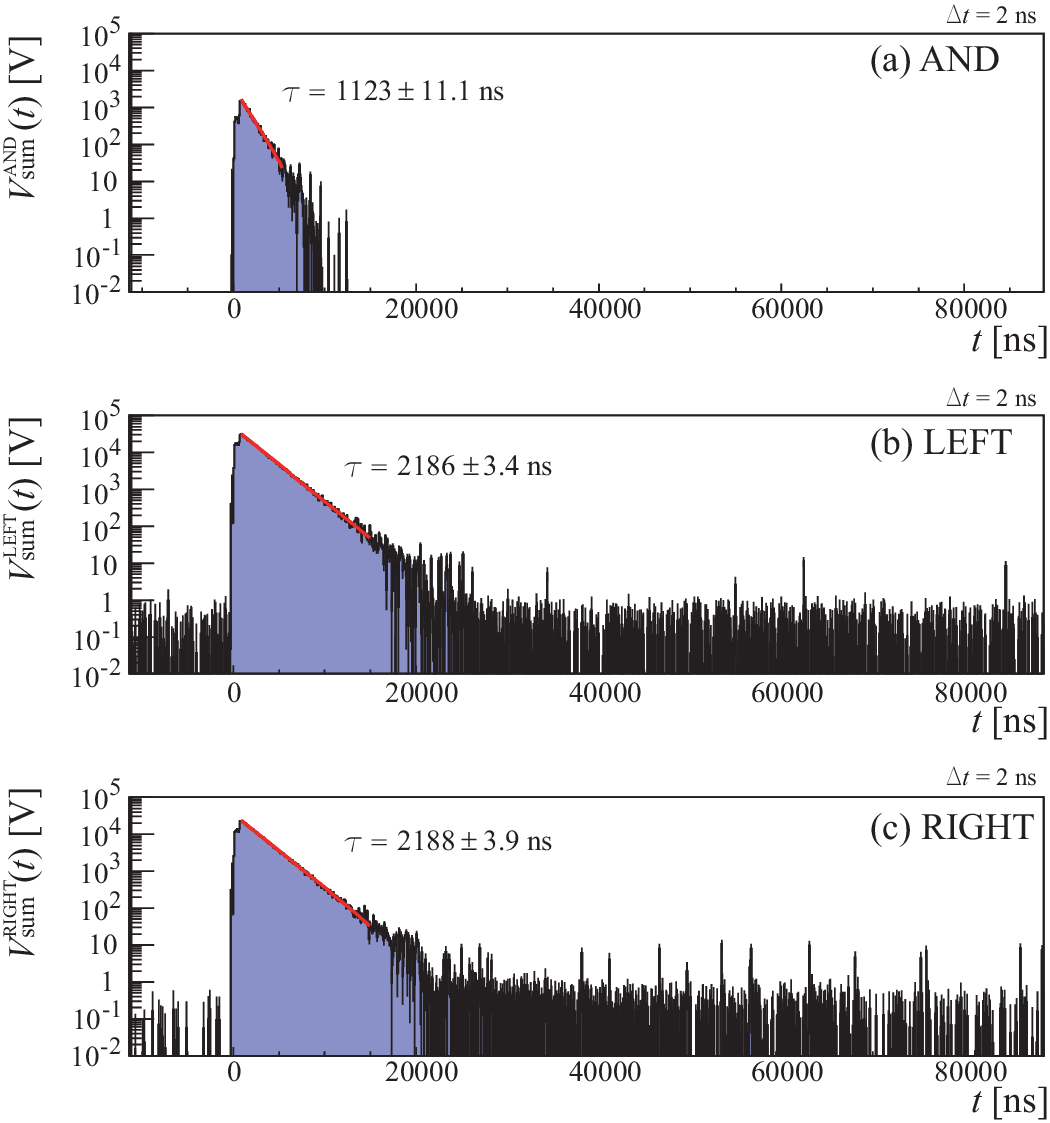}  % generate using epsconv.
 \end{center}
 \caption{
$V_{\rm sum}(t)$ spectrum for (a) LEFT $\times$ RIGHT, (b) LEFT, and (c) RIGHT. The spectrum shown in (a) is the sum of the spectral product $\sum (V^{\rm LEFT}(t) \times V^{\rm RIGHT}(t))$ and not the product of the summed spectra $V^{\rm LEFT}_{\rm sum}(t) \times V^{\rm RIGHT}_{\rm sum}(t)$. Fitted results obtained using single exponential functions are shown as solid lines.
}
 \label{and}
\end{figure}

To demonstrate this idea, we attempted to coincide the spectra of two different counters. Fig.\ref{and} presents the $V_{\rm sum}(t)$ profiles of the (a) LEFT $\cap$ RIGHT, (b) LEFT, and (c) RIGHT counters, using the data obtained through $\mu$LV-Run0. The profiles in Figs.\ref{and}(b,c) are similar to the profile in Fig.\ref{decay-0G-ch1}, showing a single exponential decay curve. Numerous background events are also apparent at $t>30,000$ ns, which are not related to muon decays and must be rejected by coinciding with the outputs of two detectors. Indeed, Kalliope has two detector layers dedicated to this purpose, but their concidence is not requested in Figs.\ref{and}(b,c).

Given that the LEFT and RIGHT counters are located opposite to the muon stopper, no real coincidence events can simultaneously hit both the LEFT and RIGHT counters. Thus, the expected result of their coincidence must be an accidental coincidence.

Fig.\ref{and}(a) presents $V_{\rm AND}(t)$ between the LEFT and RIGHT counters. The background signals shown in Figs.\ref{and}(b,c) are no longer apparent, agreeing with the observations of conventional coincidence measurements. This result demonstrates that the current-readout method can perform coincidence measurements if the rate is not too high, revealing major pileup events.

In the high-flux region, a decay curve was observed, as shown in Fig.\ref{and}(a). For this curve, the decay time constant was $\tau_{\rm AND}\sim\tau_{\mu}/2$. Mathematically, this is self-explanatory because 
$V_{\rm AND}(t)=V_{\rm LEFT}(t)\times V_{\rm RIGHT}(t)\propto e^{-t/\tau_{\mu}}\times e^{-t/\tau_{\mu}}=e^{-t/(\tau_{\mu}/2)}$. 
Nevertheless, understanding this behavior based on the probability of accidental coincidence is interesting.

Suppose there are positron events in which probability is proportional to the decay rate $p_1\propto e^{-t/\tau_{\mu}}$. 
In this case, the accidental coincidence rate becomes proportional to the event rate $p_2$ of the other counter. The resulting coincidence rate is then proportional to 
$p_1 \times p_2 \propto  e^{-t/\tau_{\mu}} \times e^{-t/\tau_{\mu}} = e^{-t/(\tau_{\mu}/2)}$. 
The obtained results corroborate this interpretation. We must emphasize that the result shown in Fig.\ref{and}(a) does not correspond to histogram resulting from the product of two histograms, $V_{\rm sum}^{\rm LEFT} \times V_{\rm sum}^{\rm RIGHT}$, but rather manifests as an over-pulse summation of $V^{\rm LEFT}(t) \times V^{\rm RIGHT}(t)$. Therefore, the result does not simply demonstrate the product of two exponential functions but indicates that the voltage product can operate as a logical AND function. This implies that we can perform coincidence measurements by applying a product operation to the two $V(t)$ time-sequential data.

Beyond radiation measurements, the application of the developed technique to logic operations could be valuable in general logic circuits. Notably, conventional two-adic Boolean algebra is suitable for strict calculations. However, to increase the data size per unit time in digital logic circuits, including computers and networks, we must increase the bus number (bus density) and/or decrease the pulse width by increasing the clock frequency.

The perspective demonstrates another possibility to increase the data size density per time and space. Beyond binary logic signals, we can extend to voltage dimensions as $n$-adic integers (or real numbers) to represent TRUENESS.

\begin{figure}[htbp]
 \begin{center}
  \includegraphics[width=86mm]{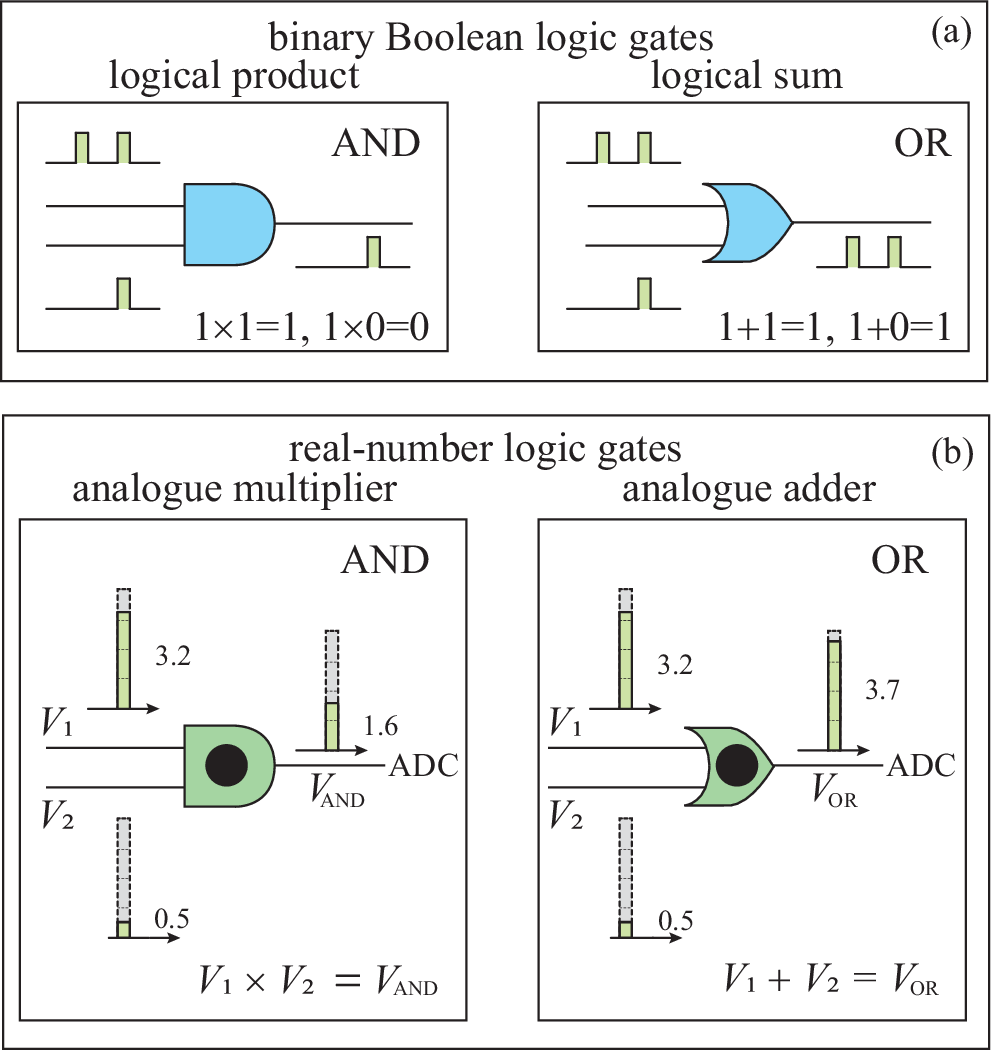} 
 \end{center}
 \caption{
AND/OR logic gates produced using analog multipliers/adders as the real-number logic gates. (a) Conventional binary logic gates. (b) Real-number logic-gates. Voltage is regarded as a representation of grayscale TRUENESS.
}
 \label{logic}
\end{figure}

Circuits similar to those in Fig.\ref{logic} can be constructed by replacing conventional AND (OR) gates with analog multiplying (summing) circuits. If the bus density and clock frequency reach their technical upper limits, this voltage TRUNESS idea may offer a solution.

\begin{figure}[htbp]
 \begin{center}
  \includegraphics[width=86mm]{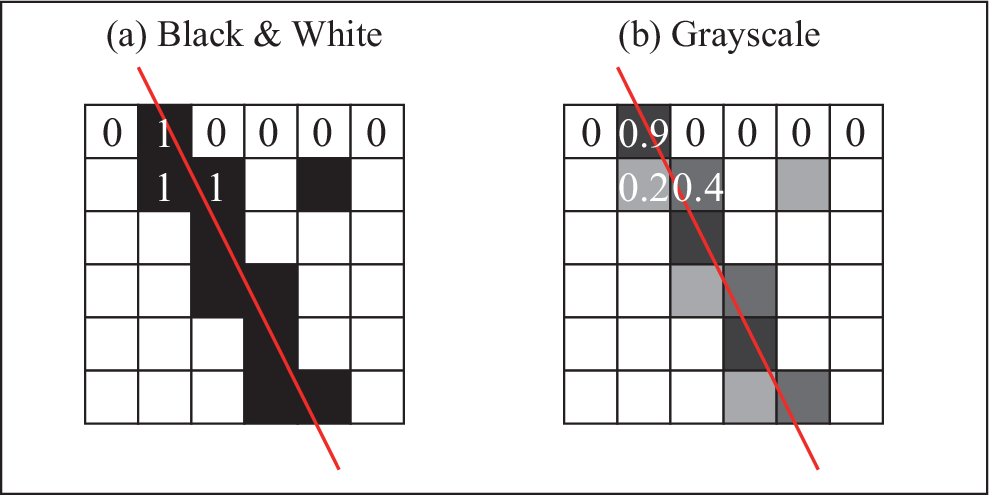}  % generate using epsconv.
 \end{center}
 \caption{
Particle tracking during radiation measurements. (a) Conventional track reconstruction using hit or no-hit events, i.e., binary logic. (b) Generalized track reconstruction incorporating the reliability of the hit in each detector channel, corresponding to the voltage value (TRUENESS) shown in Fig.\ref{logic}.
}.
 \label{tracking}
\end{figure}

This idea can also be applied to event tracking during radiation measurements. Instead of using black-and-white image data to fit a reconstruction curve, per usual practices (Fig.\ref{tracking}(a)), direct grayscale image (Fig.\ref{tracking}(b)) can be used. In that case, the image intensity represents the TRUENESS, which must be treated as the weight during fitting.

\section{Conclusion}
This study developed a new current-readout technique to handle ultra-high-counting-rate measurements, successfully confirming its application to experiments involving count rates of over 1 Gcps. 
We confirmed that the summed voltage data obtained using a digitizer represents a conventional counting histogram.
The idea was tested at J-PARC using muon decay experiments, showing it's performance.
%We also developed an approach for random uncertainty estimation, confirming the square root property, which enables us to perform $\chi^2$-analysis.
We also developed an approach for random uncertainty estimation, confirming that the random fluctuations of the summed voltage are proportional to the square root of them.
We directly confirmed that the $\chi^2$ distribution was consistent with the theoretical distribution, enabling us to perform $\chi^2$-analysis.
We also found a technique to achieve coincidence measurements using this method, which can be used as logic gates in analog logic circuits.

%%%%%%%%%%%%%%%%%%%%%%%%%%%%%%%%%%%%%%%%%%%%%%%%%%%%%%%%%%%%%%%%%%%%%%%%%%%%%%%%%%%%%%
%%%%%%%%%%%%%%%%%%%%%%%%%%%%%%%%%%%%%%%%%%%%%%%%%%%%%%%%%%%%%%%%%%%%%%%%%%%%%%%%%%%%%%

\section*{Acknowledgements}
The $\mu$LV experiments at the Materials and Life Science Experimental Facility of the J-PARC were performed under the user program 2021B0324 ($\mu$LV-Run0) and 2022B0235 ($\mu$LV- Run1). 
We thank Naritoshi Kawamura, Shoichiro Nishimura, Koichiro Shimomura, Soshi Takeshita,  Izumi Umegaki, and Takayuki Yamazaki for the detector and beamline usage.
Kenji Kojima provided detailed information required to analyze the Kalliope data.
Tomoki Harada, Shun Iimura, Hikaru Imai, Ran Ito, Kazuyoshi Kurita, Keisuke Nakamura, Hiroki Sato, Shinya Yamada, and Simon Zeidler assisted in the $\mu$LV-Run0, the Run1 or the H1 experiment.

\section*{Author contributions}
Wakata developed the data-collection system and led the $\mu$LV experiments as her master's degree work. Hara performed the PMT tests and studied the principle as part of her undergraduate work. Akamatsu and Fujiie led the H1 measurement. Murata supervised these studies, developed the data-analysis code, performed the statistical analysis and coincidence studies, and prepared the manuscript as the corresponding author. Other authors contributed to preparing the J-PARC beam tests or checking the analysis.


%\bibliographystyle{ptephy}
%\bibliography{sample}
%
% once the .bbl file has been generated then place the text in your article.

%without this code before the command "\begin{thebibliography}{}" , an error will be %flagged. When the bibliography is provided as separate .bib file, then this code %should be placed above the commands "\bibliographystyle{}" and "\bibliography{}" %inside the main TeX file. 

%\begin{thebibliography}{9}
%\bibitem{1}
%J. P.~Blaizot, and E.~Iancu, Phys. Rep. {\bf 359}, 355 (2002).
%\doi{https://doi.org/10.1016/S0370-1573(01)00061-8}
%\end{thebibliography}
\bibliography{murata}

%%%%%%%%%%%%%%%%%%%%%%%%%%%%%%%%%%%%%%%%%%%%%%%%%%%%%%%%%%%%%%%%%%%%%%%%%%%%%%%%%%%%%%

\appendix
\section{Beam intensity estimation}
\label{beam}
We estimated the beam intensity, $I_{\rm beam}$, and Kalliope's detection efficiency, $\epsilon$, using the following relations.
\begin{eqnarray}
n_{\rm single}&=&I_{\rm beam} \Omega_{\rm Kalliope}  N_{\rm ch}^{\rm single} \epsilon_{\rm deadtime}^{\rm single} \epsilon \nonumber \\
n_{\rm double}&=&I_{\rm beam} \Omega_{\rm Kalliope}  N_{\rm ch}^{\rm double} \epsilon_{\rm deadtime}^{\rm double} \epsilon^2.
\label{I-beam}
\end{eqnarray}
Here, $n_{\rm single}$ denotes the hit rate of the total 1280 channels, and $n_{\rm double}$ represents the coincidence hit rate between two layers of Kalliope, which features a combined solid angle of $\Omega_{\rm Kalliope}=0.16$ for 640 pairs and $\Omega_{\rm ch}=2.5\times 10^{-4}$ for a single pair. $\epsilon$ is defined as the detection efficiency for a single channel. Furthermore, $\epsilon_{\rm deadtime}=n_{\rm accept}/n_{\rm request}$ represents a counting loss correction factor derived from the deadtime $d$, separately estimated for single and double counts. The values of $d$ were obtained as detailed in Section \ref{saturation}.

The total numbers of channels corresponding to single and coincidence hit events were $N_{\rm ch}^{\rm single}=1280$ and $N_{\rm ch}^{\rm double}=640$, respectively. $\epsilon$ was estimated by comparing $n_{\rm single}$ and $n_{\rm double}$ as $\epsilon=n_{\rm double}/n_{\rm single}\times N_{\rm ch}^{\rm single}/N_{\rm ch}^{\rm double}=0.57$, using data corresponding to the NORMAL-intensity beam, with a minimal deadtime effect, allowing the assumption of  $\epsilon_{\rm dead \; time}^{\rm double}\sim 1$. This assumption was made because the decay curve recorded by Kalliope under NORMAL-intensity exposure demonstrated negligible saturation.  $\epsilon_{\rm dead \; time}^{\rm double} \sim 0.7$ (for $t=0$ timing) was obtained for the data corresponding to the FULL-intensity beam.
Using Eq.(\ref{I-beam}) with the above numbers, the $I_{\rm beam}$ values for the NORMAL and FULL beam were obtained.

%\section{Principle and statistics}
%\subsection{Electric Circuits}
\section{Electric Circuits}
%If we are interested in investigating relatively slow phenomena with a timescale on the order of  $\upmu$s, as in the current study, a time resolution of  $\sigma_t\sim100 \;{\rm ns}$, for instance, is sufficient to observe the time spectra of  $n_e(t)$. In such cases, identifying discrete pulse signals with ns precision, as required in conventional event-by-event pulse counting, is unnecessary. Instead, obtaining their averaged flux (event rate) information (i.e., current) as a function of time $t$ with a resolution of $\sigma_t$ is adequate. Therefore, in this study, we attempted to obtain flux information from the output electric currents  $i(t)$ of the PMTs.

\begin{figure}[htbp]
 \begin{center}
  \includegraphics[width=86mm]{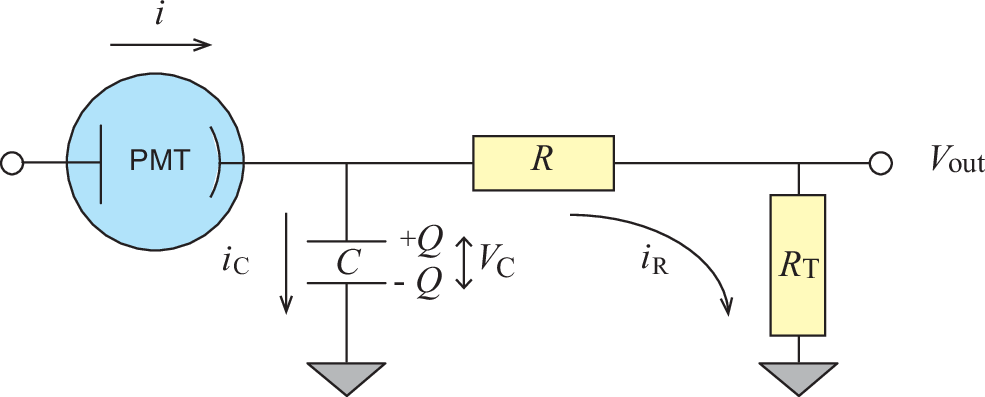} 
 \end{center}
 \caption{
Equivalent electric circuit for PMT readout. Here, the PMT acts as a current source driving $i$, and $V_{\rm out}$ denotes the voltage between termination resistor $R_{\rm T}$ to be measured. $R$ is the series resistance of the signal cable, and $C$ is the total capacitance of the cable and the PMT. $\pm Q$ is charge on the capacitor, $V_{\rm C}$ is the voltage between them. $i_{\rm C}$ and $i_{\rm R}$ are the current flew though the capacitor and resistors, respectively.}
%Equivalent electric circuit for PMT readout. Here, the PMT acts as a current source, and $V_{\rm out}$ denotes the voltage to be measured.}
 \label{circuit}
\end{figure}

Fig.\ref{circuit} shows the equivalent electric circuit for the expected voltage, $V_{\rm out}$, considering that a PMT is a current source \cite{Knoll}. The termination resistance, $R_{\rm T}$, satisfies the following differential equation:
\begin{equation}
i(t)=C\;\frac{R+R_{\rm T}}{R_{\rm T}} 
\frac{dV_{\rm out}(t)}{dt}
+\frac{V_{\rm out}(t)}{R_{\rm T}}.
\label{diff}
\end{equation}
The above relation is derived based on the following:
\begin{eqnarray}
    &&i=i_{\rm C}+i_{\rm R}, \nonumber\\
    &&i_{\rm C}=\frac{dQ}{dt}=C\frac{dV_{\rm C}}{dt},\nonumber\\
    &&V_{\rm C}=V_{\rm out}+i_{\rm R}R,\nonumber\\
    &&V_{\rm out}=i_{\rm R} R_{\rm T},
\end{eqnarray}
where $C$ denotes the total capacitance of the PMT and the signal cable, with a typical value of $\sim 100 \,{\rm pF}$. Given that the series resistance, $R$, of the cable must be extremely small, we can assume that $R/R_{\rm T}\ll1$ in usual cases. Thus, Eq.(\ref{diff}) simplifies to
\begin{equation}
i(t)\sim C
\frac{dV_{\rm out}(t)}{dt}
+\frac{V_{\rm out}(t)}{R_{\rm T}}.
\label{diff2}
\end{equation}
The characteristics of the solution of Eq.(\ref{diff2}) are well known, as described in Knoll's textbook \cite{Knoll}. If the input current $i(t)$ exhibits a narrow pulse structure, the solution, $V_{\rm out}(t)$, is anticipated to display a pulse shape with a rising time corresponding to the width of the input current. A time constant $\tau_{\rm c}=R_{\rm T}C$ represents the exponential falling time. %Accordingly, $V_{\rm out}(t)$, read between the digitizer's input resistor $R_{\rm T}$, represents the output current $i(t)$, whose integrated value is proportional to the number of photons produced within the scintillator.
Accordingly, $V_{\rm out}(t)$, read between the digitizer's input resistor $R_{\rm T}$, represents the output current $i(t)$, whose integrated value is proportional to the number of photons produced within the scintillator, supposing a constant amplification factor of the PMT.

\begin{figure}[htbp]
 \begin{center}
  \includegraphics[width=75mm]{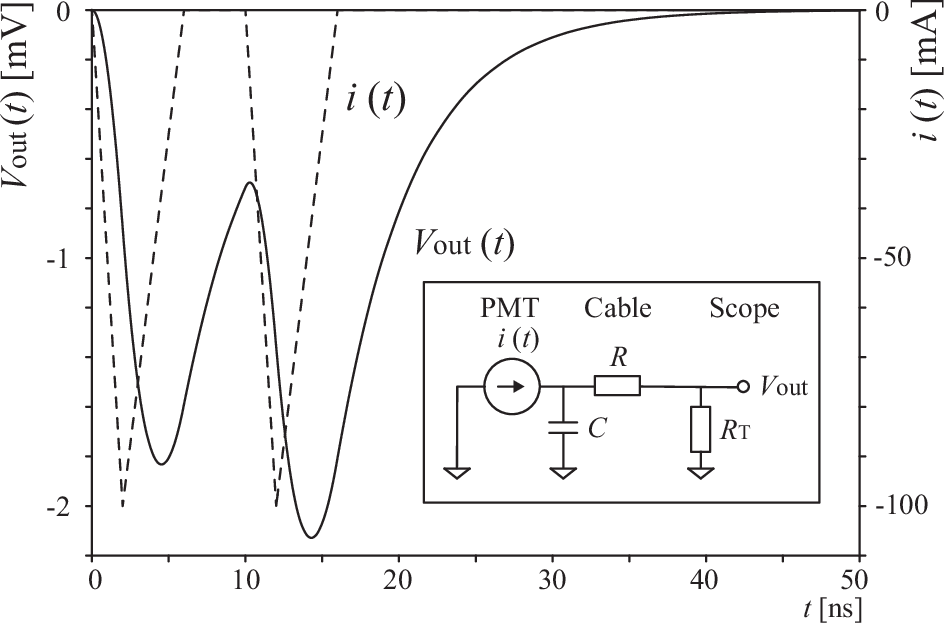} 
 \end{center}
 \caption{
Results of the circuit simulation conducted using LTspice, assuming parameter values of $C=10 $ pF, $R_{\rm T}=50\;\Omega$, and $R=0.1 \; \Omega$ and assuming that $i(t)$ generates $V_{\rm out}(t)$ \cite{LTspice}.}
 \label{spice}
\end{figure}

This estimation was confirmed by performing a circuit simulation (LTspice \cite{LTspice}), and the results of this simulation are shown in Fig.\ref{spice}. In this simulation, the input signal comprised a pair of closely spaced triangular current pulses, as illustrated in the figure. As anticipated, the output voltage signals, $V_{\rm out}(t)$, were nearly proportional to the input current, $i(t)$, with a decay time constant of $\tau_{\rm c}$. This result demonstrates the pileup phenomenon between two output voltage pulses. To overcome this bottleneck, we must acknowledge that such pileup data contains two independent events.

%The quantity to be measured is the number of positrons hitting the detector, rather than the number of photons generated by scintillators. The number of photons corresponding to a single positron penetration event typically fluctuates owing to the changing path lengths of positrons penetrating the scintillators, statistical variations in the number of photons, and statistical fluctuations in the number of secondary electrons generated within the PMTs. These complex pulse-to-pulse charge fluctuations can be avoided through discrimination when applying the conventional pulse-counting method. On the other hand, these analog fluctuations are anticipated to influence the final precision of the flux measurement in the present method. Therefore, the final precision for $n_e(t)$ may be worse than that in the conventional pulse-counting method; consequently, we must quantitatively estimate the contributions of these additional fluctuations, as discussed in Section \ref{voltage-fluctuation}.

%%%%%%%%%%%%%%%%%%%%%%%%%%%%%%%%%%%%%%%%%%%%%%%%%%%%%%%%%%%%%%%%%%%%%%%%%%%%%%%%%%%%%%
%%%%%%%%%%%%%%%%%%%%%%%%%%%%%%%%%%%%%%%%%%%%%%%%%%%%%%%%%%%%%%%%%%%%%%%%%%%%%%%%%%%%%%

%%%%%%%%%%%%%%%%%%%%%%%%%%%%%%%%%%%%%%%%%%%%%%%%%%%%%%%%%%%%%%%%

\end{document}